\documentclass[journal]{IEEEtran}

\usepackage{graphicx}

\usepackage{bm}
\usepackage{amsfonts}

\usepackage{amssymb}

\usepackage{color}

\usepackage{cite}

\ifCLASSINFOpdf
\else
\fi
\usepackage{amsmath}
\usepackage{algorithm}
\usepackage{algorithmic}

\usepackage{booktabs}

\ifCLASSOPTIONcompsoc
 \usepackage[caption=false,font=normalsize,labelfont=sf,textfont=sf]{subfig}
\else
 \usepackage[caption=false,font=footnotesize]{subfig}
\fi

\usepackage{stfloats}
\fnbelowfloat
\begin{document}
%
\title{MU-MIMO Communications with MIMO Radar: From Co-existence to Joint Transmission}
%
%

\author{Fan Liu,~\IEEEmembership{Student~Member,~IEEE,}
        Christos Masouros,~\IEEEmembership{Senior~Member,~IEEE,}
        Ang Li,~\IEEEmembership{Student~Member,~IEEE,}
        Huafei Sun
        and~Lajos Hanzo,~\IEEEmembership{Fellow,~IEEE}
\thanks{F. Liu is with the School of Information and Electronics, Beijing Institute of Technology, Beijing, 100081, China, and is also with the Department of Electronic and Electrical Engineering, University College London, London, WC1E 7JE, UK. (e-mail: liufan92@bit.edu.cn).}
\thanks{C. Masouros and A. Li are with the Department of Electronic and Electrical Engineering, University College London, London, WC1E 7JE, UK. (e-mail: chris.masouros@ieee.org, ang.li.14@ucl.ac.uk).}
\thanks{H. Sun is with the School of Mathematics and Statistics, Beijing Institute of Technology, Beijing, 100081, China. (email: huafeisun@bit.edu.cn).}
\thanks{L. Hanzo is with the School of Electronics and Computer Science, University of Southampton, Southampton SO17 1BJ, UK. (e-mail: lh@ecs.soton.ac.uk).}
\thanks{This work has been submitted to the IEEE for possible publication.  Copyright may be transferred without notice, after which this version may no longer be accessible.}
}

\maketitle
\begin{abstract}
Beamforming techniques are proposed for a joint multi-input-multi-output (MIMO) radar-communication (RadCom) system, where a single device acts both as a radar and a communication base station (BS) by simultaneously communicating with downlink users and detecting radar targets. Two operational options are considered, where we first split the antennas into two groups, one for radar and the other for communication. Under this deployment, the radar signal is designed to fall into the null-space of the downlink channel. The communication beamformer is optimized such that the beampattern obtained matches the radar's beampattern while satisfying the communication performance requirements. To reduce the optimizations' constraints, we consider a second operational option, where all the antennas transmit a joint waveform that is shared by both radar and communications. In this case, we formulate an appropriate probing beampattern, while guaranteeing the performance of the downlink communications. By incorporating the SINR constraints into objective functions as penalty terms, we further simplify the original beamforming designs to weighted optimizations, and solve them by efficient manifold algorithms. Numerical results show that the shared deployment outperforms the separated case significantly, and the proposed weighted optimizations achieve a similar performance to the original optimizations, despite their significantly lower computational complexity.
\end{abstract}


%

\section{Introduction}
%
%
%
%
\IEEEPARstart{F}{ree} frequency spectrum has become a valuable asset. It has been reported that by 2020, the number of connected devices will exceed 20 billion, which requires extra spectrum resources. Below 10GHz, the S-band (2-4GHz) and C-band (4-8GHz) are occupied by a variety of radar applications at the time of writing, and it is envisaged that these can be shared with communication systems in the future \cite{federal2010connecting,ssparc,specees}. While policy and regulations may delay the practical application of such solutions, research efforts are well under way to address the practical implementation of communication and radar spectrum sharing (CRSS). Existing contributions on CRSS mainly focus on two aspects: 1) co-existence of existing radar and communication devices and 2) co-design for dual-functional systems. Below we review the related literature in both cases.
\subsection{Co-existence of Radar and Communications Systems}
As a straightforward way to support the co-existence, opportunistic spectrum sharing between rotating radar and cellular systems has been considered in \cite{6331681}, where the communication system transmits its signals, when the space and frequency spectra are not occupied by radar. While such a scheme seems easy to realize, it does not allow the radar and communication to work simultaneously. Recognizing this fact, the pioneering work of Sodagari et al. \cite{6503914} proposes a null-space projection (NSP) method to support the co-existence of MIMO radar and BS, in which a MIMO radar beamformer is designed to project the radar signals onto the null-space of the interference channels between the radar and BS, thus zero-forcing the interference imposing on the communication link. However, such a beamformer may result in performance loss for the radar. Further contributions \cite{7814210,7089157,7289385} investigate different trade-offs between the performance of radar and communications by relaxing the zero-forcing precoder as the projection matrix generated by the singular vectors, whose singular values are less than certain thresholds, and therefore imposes controllable interference levels on the communication systems.
\\\indent Recent treatises \cite{7485158,7470514,7953658,7511108,7898445,liu2017interference} have employed sophisticated optimization techniques to solve the problem. In \cite{7485158}, the radar beamformer and communication covariance matrix are jointly designed to maximize the radar's SINR subject to specific capacity and power constraints. Similar work has been done in \cite{7470514,7953658} for the co-existence between the MIMO matrix completion (MIMO-MC) radar and point-to-point (P2P) MIMO communications, where the radar sub-sampling matrix is further introduced as a variable in the optimizations. To address the co-existence of MIMO radar and multi-user MIMO (MU-MIMO) communications, Liu et al. \cite{7898445} considers the issues of robust beamforming design for the BS with imperfect channel state information (CSI), where the radar detection probability is maximized subject to satisfying the SINR requirements of downlink users and the power budget of the BS. As a further step, a novel CRSS beamforming optimization is proposed by Liu et al. \cite{liu2017interference} by exploiting the multi-user interference (MUI) as a useful source of transmission power, which leads to significant power-saving compared to the conventional methods.

\subsection{Co-design of Dual-functional RadCom Systems}
It is worth noting that for supporting co-existence, radar and communication devices are typically required to exchange side-information for achieving a beneficial cooperation, such as the CSI, radar probing waveforms, communication modulation format and frame structure. In practical scenarios, however, these requirements are difficult to implement. Hence, a more favorable approach for CRSS is to design a novel dual-functional system that carries out both radar and communications\cite{7782415}, where the above problem does not exist. Aiming for finding the fundamental performance bounds of a joint RadCom system, an information theoretical framework has been presented in \cite{7279172,7855671} to unify radar and communications with the help of rate distortion theory. Similar efforts have been made by Guerci et al. \cite{7131098}, where the concept of radar information rate has been defined for ensuring that the performance of the two functions can be discussed under similar metrics. In addition, waveform design is also considered as an enabling solution, where a dual-functional waveform is designed to supporting both cellular communication and target detection at the same time. Existing contributions on single-antenna systems study the feasibility of time-division waveforms \cite{7409935}, the combination of the communication symbols with the radar waveform carrier \cite{4268440}, and the use of communication waveforms for radar detection \cite{5776640}. Nevertheless, all of these proposals lead to significant performance losses imposed on either the radar or the communications side. More recently, waveform design considered for MIMO systems has drawn much attention from researchers, where the information bits are modulated by controlling the sidelobe levels of the transmit beampattern of the MIMO radar \cite{7347464}, or by Phase Shift Keying (PSK) modulation via waveform diversity  \cite{7575457}. While such schemes are well-designed, the transmission rate obtainable for communication remains relatively low, since the communication symbol is embedded in several radar waveforms.

\subsection{The Contribution of Our Work}
In this paper, we develop a series of optimization-based transmit beamforming approaches for a joint MIMO RadCom system, which is defined as a dual-functional platform that can simultaneously transmit probing signals to radar targets and serve multiple downlink users. Such a system realizes both functions within the same frequency band, with significant practical applications. Unlike the conventional joint MIMO waveform design approaches using sidelobe control or waveform diversity, the proposed methods consider the beamforming matrices rather than the waveforms, which do not affect the orginal modulation scheme of the communication. We note that for such systems, the designs of the radar receiver as well as of the uplink communications are also challenging problems. Due to the limited space, and following the approaches of \cite{7347464,7575457}, we only focus our attention on the joint transmission of the RadCom system in this paper, and set aside the receiver techniques for our future work.
\\\indent Inspired by the NSP method in \cite{6503914,7814210,7089157,7289385}, we firstly consider a separated antenna deployment, where the antennas are partitioned into radar and communication antennas. The interference imposed by the radar signals on the users are eliminated by zero-forcing, and the communication covariance matrix is designed to formulate an appropriate beampattern that matches the radar beampattern, while satisfying the communication performance requirements. The zero-forcing beamforming problem is non-convex, but can be solved by the classic semidefinite relaxation (SDR) technique.
\\\indent To alleviate the optimization constraints imposed by the above mentioned separated design, we further propose a shared deployment, where all the antennas are shared for both radar detection and downlink communications. This implies that the communication signals are also used as radar probing waveforms. Aiming for designing an appropriate radar beampattern, while satisfying the downlink power and SINR constraints, a non-convex joint optimization problem is formulated, which can be also solved via SDR.
\\\indent To further simplify the problem, we design the corresponding weighted optimizations by incorporating SINR constraints as penalty terms in the objective functions. We consider a pair of different penalty terms as well as two power constraints, i.e. the sum-squared SINR error penalty, the max SINR penalty, the total power constraint and the per-antenna power constraint. By doing so, the feasible regions of the weighted problems are viewed as Riemannian manifolds \cite{petersen2006riemannian}, and therefore can be solved by manifold optimization algorithms \cite{absil2009optimization}. It has been shown that these algorithms are more computationally efficient for the non-convex precoding optimizations with equality constraints than conventional methods, and yield near-optimal solutions without approximations\cite{7898517,7811286,liu2017CEP}. Recent treatises also investigate the application of such algorithms on control theory \cite{duan2013natural,li2016optimal}, where a fast convergence rate can be achieved. Finally, the computational complexities of the proposed algorithms are analyzed in terms of floating-point operations (flops). For clarity, we list the contributions of this paper as follows:
\begin{itemize}
\item We propose both separated and shared antenna deployments for the joint RadCom system, and consider the corresponding beamforming design problems.
\item We design a number of weighted beamforming optimizations for the shared antenna deployment in conjunction with different constraints and penalty terms.
\item We reformulate the weighted optimization problems as manifold optimizations, which achieve an identical performance to that of the original optimization by using low-complexity manifold solvers.
\item We derive the computational complexity analytically in terms of the number of the flops when using the proposed algorithm to solve weighted optimizations.
\end{itemize}

The remainder of this paper is organized as follows, Section II introduces the system model, Section III formulates the optimization problems, Section IV  solves the beamforming problems using manifold based algorithms, Section V analyzes the computational complexity of the proposed algorithms, Section VI provides numerical results, and Section VII concludes the paper.
\\\indent \emph{Notations}: Unless otherwise specified, matrices are denoted by bold uppercase letters (i.e., ${\bf{H}}$), bold lowercase letters are used for vectors (i.e., $\pmb\alpha$), scalars are denoted by normal font (i.e., $\gamma_i$), subscripts indicate the columns of a matrix (i.e., ${\bf{h}}_i$ is the \emph{i}-th column of ${\bf{H}}$), $\operatorname{tr}\left(\cdot\right)$ stands for the trace of the argument, $(\cdot)^T$, $(\cdot)^*$ and $(\cdot)^H$ stand for transpose, complex conjugate and Hermitian transpose respectively, $\operatorname{Re}(\cdot)$ denotes the real part of the argument, $\operatorname{diag}(\cdot)$ stands for the vector composed by the diagonal entries of a matrix, $\operatorname{ddiag}(\cdot)$ sets all off-diagonal entries of a matrix to zero, $\left\| \cdot\right\|$ and $\left\| \cdot\right\|_F$ denote the $l_2$ norm and the Frobenius norm respectively.


\section{System Model}

\begin{figure*}[!t]
\centering
\subfloat[]{\includegraphics[width=0.8\columnwidth]{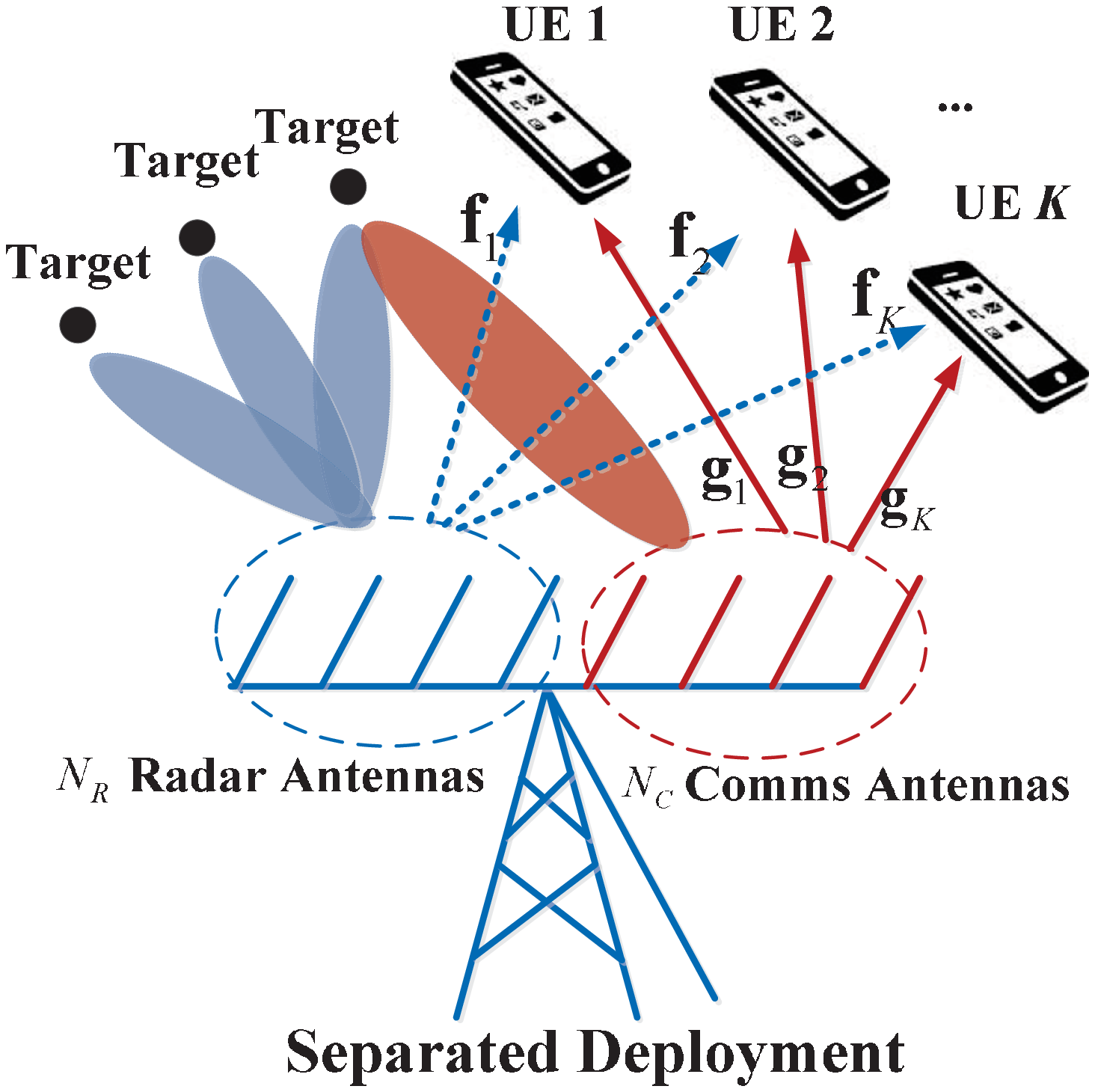}
\label{fig.1}}
\hspace{0.1in}
\subfloat[]{\includegraphics[width=0.8\columnwidth]{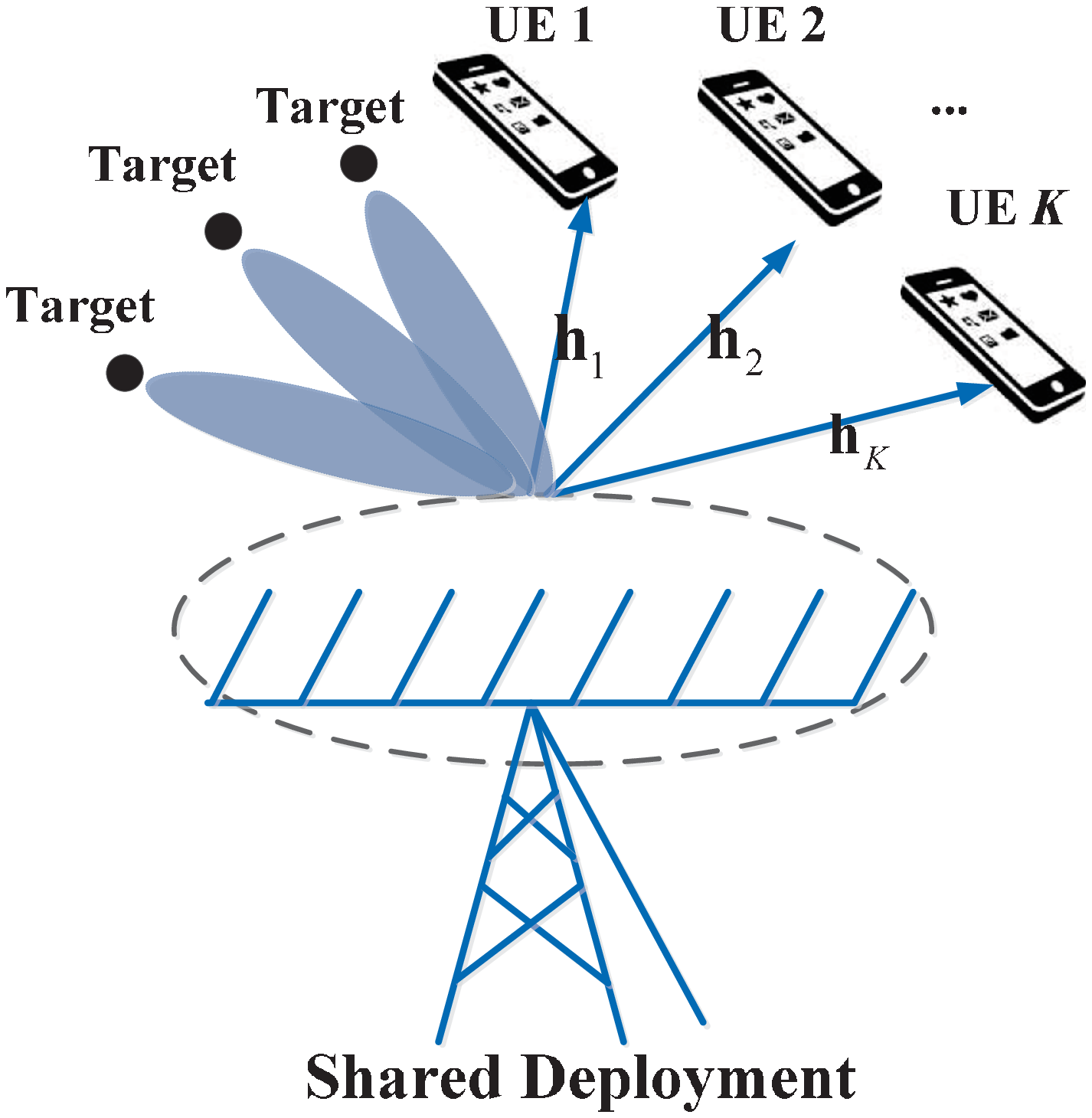}
\label{fig.2}}
\caption{Antenna deployments for the collocated radar and communications BS. (a) Separated deployment; (b) Shared deployment.}
\label{fig_system}
\end{figure*}

We consider a joint MIMO RadCom system, which can simultaneously transmit probing signals to the targets located at the angles of interest and communication symbols to downlink users. The joint system is equipped with a uniform linear array (ULA) of \emph{N} antennas, serving \emph{K} single-antenna users in the downlink while detecting targets at the same time. Below we present the signal models for the pair of operations considered in this paper, namely the separated deployment and the shared deployment.
\subsection{Separated Deployment}
As shown in Fig. 1 (a), the separated deployment involves splitting the antenna array into two groups: one for radar and one for the downlink communications. In this case, the received signal of the \emph{i}-th user is given by
\begin{equation}
{x_i}[l] = {\mathbf{g}}_i^T\sum\limits_{k = 1}^K {{{\mathbf{w}}_k}{d_k}[l]}  + {\mathbf{f}}_i^T{{\mathbf{s}}_l} + {\omega_i}[l],\forall i,
\end{equation}
where ${{\mathbf{g}}_i}\in {\mathbb{C}^{N_C \times 1}},{{\mathbf{f}}_i} \in {\mathbb{C}^{N_R\times 1}}$ are the channel vectors from communication antennas and radar antennas to the \emph{i}-th user, $N_C$ and $N_R$ are the number of antennas dedicated to communication and radar respectively, $ {d}_i[l] $ and $ {\omega}_i[l] \sim {\mathcal{C}}{\mathcal{N}}\left( {0,N_0} \right)$ stand for the communication symbol and the received noise of the \emph{i}-th user at the time index \emph{l}, $ {\bf{w}}_{i} \in {\mathbb{C}^{N_C \times 1}} $ denotes the beamforming vector of the \emph{i}-th user, and finally ${{\mathbf{s}}_l} \in {\mathbb{C}^{N_R \times 1}}$ is the \emph{l}-th snapshot across the radar antennas. The sample covariance matrix of the radar signal is given as $\frac{1}{L}\sum\limits_{l = 1}^L {{{\mathbf{s}}_l}{\mathbf{s}}_l^H}  = {\mathbf{R}}_1 \in {\mathbb{C}^{N_R \times N_R}}$, with $ L $ being the length of the signal on the fast-time axis. We rely on the following standard assumptions:
\begin{enumerate}
\item The communication signals are statistically independent of the radar signals;
\item The channel between the RadCom system and users is flat Rayleigh fading, which is given by $ {\bf{H}}=\left[{\bf{h}}_1, {\bf{h}}_2,...,{\bf{h}}_K \right] $ , where ${{\mathbf{h}}_i} = \left[ {{{\mathbf{f}}_i};{{\mathbf{g}}_i}} \right]\in {\mathbb{C}^{\emph{N} \times 1}}$. It is assumed that the channel is perfectly estimated by pilot symbols.
\end{enumerate}
Note that the separated deployment allows \emph{arbitrary} radar signals to be used.
\\\indent By letting ${\mathbf{W}}_k={{\bf{w}}_k}{{\bf{w}}_k^H}$, the communication transmit power is given as
\begin{equation}
    {P_1} = \sum\limits_{k = 1}^K {{{\left\| {{{\bf{w}}_k}} \right\|}^2}} = \sum\limits_{k = 1}^K {\operatorname{tr} \left( {{{\mathbf{W}}_k}} \right)}.
\end{equation}
Based on (1), the received SINR of the \emph{i}-th user is thus given by
\begin{equation}
\begin{gathered}
  {\beta _i} = \frac{{{{\left| {{\mathbf{g}}_i^T{{\mathbf{w}}_i}} \right|}^2}}}{{\sum\limits_{\begin{subarray}{l}
  k = 1 \\
  k \ne i
\end{subarray}}^K {{{\left| {{\mathbf{g}}_i^T{{\mathbf{w}}_k}} \right|}^2}}  + {\mathbf{f}}_i^T{{\mathbf{R}}_1}{\mathbf{f}}_i^* + {N_0}}}\; \hfill \\
  \;\;\;\; = \frac{{\operatorname{tr} \left( {{\mathbf{g}}_i^*{\mathbf{g}}_i^T{{\mathbf{W}}_i}} \right)}}{{\operatorname{tr} \left( {{\mathbf{g}}_i^*{\mathbf{g}}_i^T\sum\limits_{\begin{subarray}{l}
  k = 1 \\
  k \ne i
\end{subarray}}^K {{{\mathbf{W}}_k}} } \right) + \operatorname{tr} \left( {{\mathbf{f}}_i^*{\mathbf{f}}_i^T{{\mathbf{R}}_1}} \right) + {N_0}}}. \hfill \\
\end{gathered}
\end{equation}
The covariance matrix for the precoded communication symbols can be obtained by
\begin{equation}
{\mathbf{C}}_1  = \sum\limits_{k = 1}^K {{{\mathbf{W}}_k}} .
\end{equation}

\subsection{Shared Deployment}
In this case, all the \emph{N} antennas are shared for both radar detection and downlink communication. This concept is shown in Fig. 1 (b). The received signal at the \emph{i}-th user is given as
\begin{equation}
    {{y}_i[l]} = {\bf{h}}_i^T\sum\limits_{k = 1}^K {{{\bf{t}}_k}{d}_k[l]} + {n}_i[l], \forall i,
\end{equation}
where  $ {\bf{t}}_{i} \in {\mathbb{C}^{\emph{N} \times 1}} $ and $ {n}_i[l] \sim {\mathcal{C}}{\mathcal{N}}\left( {0,N_0} \right)$ denote the beamforming vector and the received noise of the \emph{i}-th user, respectively. We make the following assumptions:
\begin{enumerate}
\item The RadCom system employs the communication signal as the radar probing waveform;
\item As above, the channel $\mathbf{H}$ between the RadCom system and the users is Rayleigh flat fading and it is also perfectly estimated.
\end{enumerate}
At first glance, it seems that there are no degrees of freedom (DoF) in designing the radar signal, since the communication signal is employed as a dual-functional waveform in the shared deployment. Nevertheless, we will show in the following sections that this is indeed an affordable constraint, given the resultant benefits. Similarly, the transmit power is given by
\begin{equation}
    {P_2} = \sum\limits_{k = 1}^K {{{\left\| {{{\bf{t}}_k}} \right\|}^2}} = \sum\limits_{k = 1}^K {\operatorname{tr} \left( {{{\mathbf{T}}_k}} \right)},
\end{equation}
where ${\mathbf{T}}_k={{\bf{t}}_k}{{\bf{t}}_k^H}$.
The received SINR of the \emph{i}-th user is given by
\begin{equation}
{\gamma _i} = \frac{{{{\left| {{\mathbf{h}}_i^T{{\mathbf{t}}_i}} \right|}^2}}}{{\sum\limits_{\begin{subarray}{l}
  k = 1 \\
  k \ne i
\end{subarray}}^K {{{\left| {{\mathbf{h}}_i^T{{\mathbf{t}}_k}} \right|}^2}}  + {N_0}}} = \frac{{\operatorname{tr} \left( {{\mathbf{h}}_i^*{\mathbf{h}}_i^T{{\mathbf{T}}_i}} \right)}}{{\operatorname{tr} \left( {{\mathbf{h}}_i^*{\mathbf{h}}_i^T\sum\limits_{\begin{subarray}{l}
  k = 1 \\
  k \ne i
\end{subarray}}^K {{{\mathbf{T}}_k}} } \right) + {N_0}}}.{\text{ }}
\end{equation}
Note that in (7) no radar interference is imposed on the users.
The covariance matrix of the precoded symbols is given by
\begin{equation}
{\mathbf{C}}_2 = \sum\limits_{k = 1}^K {{{\mathbf{T}}_k}} .
\end{equation}

\section{Problem Formulation}
In the proposed scenario, an appropriate beamformer has to be designed to meet the following requirements:
\begin{itemize}
\item To generate a close match to the desired beampattern for radar detection;
\item To guarantee the SINR level required for the downlink users;
\item To satisfy the transmit power budget.
\end{itemize}
In this section, we first recall the beampattern designs for the colocated MIMO radar only, and then combine them with communication constraints for ensuring that the beamformer obtained can indeed meet the above criteria.

\subsection{MIMO Radar Beampattern Design}
It is widely exploited that by employing uncorrelated waveforms, MIMO radar has higher DoFs than the traditional phased-array radar \cite{li2009mimo,4276989,4350230}. The existing literature indicates that the design of such a beampattern is equivalent to designing the covariance matrix of the probing signals, where convex optimization can be employed. In \cite{4516997}, a constrained least-square problem is formulated to approach an ideal beampattern. Here we recapture it as follows
\begin{subequations}  \label{eq:9}
\begin{align}
    & \mathop {\min }\limits_{\alpha ,{\mathbf{R}}} \sum\limits_{m = 1}^M {{{\left| {\alpha {{\tilde P}_d}\left( {{\theta _m}} \right) - {{\mathbf{a}}^H}\left( {{\theta _m}} \right){\mathbf{Ra}}\left( {{\theta _m}} \right)} \right|}^2}} \label{eq:9a}  \\
    & s.t.\;\;\operatorname{diag} \left( {\mathbf{R}} \right) = \frac{{{P_0}{\mathbf{1}}}}{{{N_t}}},\label{eq:9b} \\
    & \;\;\;\;\;\;\;\;{\mathbf{R}} \succeq 0,{\mathbf{R}} = {{\mathbf{R}}^H},         \label{eq:9c} \\
    & \;\;\;\;\;\;\;\;\alpha  \ge 0, \label{eq:9d}
\end{align}
\end{subequations}
where $\left\{ {{\theta _m}} \right\}_{m = 1}^M$ is defined as a fine angular grid that covers the detection angle range of $\left[-\pi/2,\pi/2\right]$, ${\mathbf{a}}\left( \theta_m  \right) = \left[ {1,{e^{j2\pi \Delta \sin \left( \theta_m  \right)}},...,{e^{j2\pi \left( {N_t - 1} \right)\Delta \sin \left( \theta_m  \right)}}} \right]^T \in {\mathbb{C}^{N \times 1}}$ is the steering vector of the transmit antenna array with $\Delta$ being the spacing between adjacent elements normalized by the wavelength, $N_t$ is the number of antennas of the array, ${{\tilde P}_d}\left( \theta_m  \right)$ is the desired ideal beampattern gain at $\theta_m$, $\bf R  $ is the waveform covariance matrix, $P_0$ is the power budget, $\alpha$ is a scaling factor, and $\mathbf{1}$ is defined as $\mathbf{1} = \left[1,1,...,1\right]^T\in{\mathbb{R}^{N\times1}}$. The constraint (9b) is imposed to guarantee that the waveform transmitted by different antennas has the same average power.
\\\indent Aiming for generating a beampattern with a desired 3dB main-beam width, another optimization problem has been proposed by Stoica et al.\cite{4276989}, which is given by
\begin{equation}
\begin{gathered}
  \mathop {\min }\limits_{t,{\mathbf{R}}} {\kern 1pt} {\kern 1pt} {\kern 1pt}  - t \hfill \\
  s.t.\;\;\;{{\mathbf{a}}^H}\left( {{\theta _0}} \right){\mathbf{Ra}}\left( {{\theta _0}} \right) - {{\mathbf{a}}^H}\left( {{\theta _m}} \right){\mathbf{Ra}}\left( {{\theta _m}} \right) \ge t,\forall {\theta _m} \in \Omega,  \hfill \\
  \;\;\;\;\;\;\;{\kern 1pt} {\kern 1pt} {{\mathbf{a}}^H}\left( {{\theta _1}} \right){\mathbf{Ra}}\left( {{\theta _1}} \right) = {{\mathbf{a}}^H}\left( {{\theta _0}} \right){\mathbf{Ra}}\left( {{\theta _0}} \right)/2, \hfill \\
  \;\;\;\;\;\;\;\;{{\mathbf{a}}^H}\left( {{\theta _2}} \right){\mathbf{Ra}}\left( {{\theta _2}} \right) = {{\mathbf{a}}^H}\left( {{\theta _0}} \right){\mathbf{Ra}}\left( {{\theta _0}} \right)/2,\;\; \hfill \\
  \;\;\;\;\;\;\;\;{\mathbf{R}} \succeq 0,{\mathbf{R}} = {{\mathbf{R}}^H}, \hfill \\
  \;\;\;\;\;\;\;\;\operatorname{diag}\left( {\mathbf{R}} \right) = \frac{{{P_0}{\mathbf{1}}}}{N_t}, \hfill \\
\end{gathered}
\end{equation}
where $\theta_0$ is the location of the main-beam, $\left(\theta_2-\theta_1\right)$ determines the 3dB main-beam width, and $\Omega$ denotes the sidelobe region. Note that the above two problems are convex, and thus can be efficiently solved by numerical tools.
\subsection{Zero-forcing Beamforming for Separated Deployment}
We first consider the beamforming design of the separated deployment. Motivated by the NSP method, which has been widely applied to radar and communication co-existence scenarios \cite{6503914,7814210,7089157,7289385}, we force the radar signals to fall into the null-space of the channel between the radar antennas and downlink users to eliminate the interference. This can be equivalently written as
\begin{equation}
\mathbb{E}\left\{ {{{\left\| {{\mathbf{f}}_i^T{{\mathbf{s}}_l}} \right\|}^2}} \right\} = {\mathbf{f}}_i^T\mathbb{E}\left\{ {{{\mathbf{s}}_l}{\mathbf{s}}_l^H} \right\}{\mathbf{f}}_i^* = \operatorname{tr} \left( {{\mathbf{f}}_i^*{\mathbf{f}}_i^T{{\mathbf{R}}_1}} \right) = 0,\forall i.
\end{equation}
By introducing the above constraint in (9) and (10), and using $N_R$ antennas for radar detection, an $N_R \times N_R$ covariance matrix ${\bf R}_1$ can be obtained. Accordingly, the zero-forcing version for (9) and (10) are given by
\begin{equation}
\begin{array}{*{20}{l}}
  \displaystyle \mathop {\min }\limits_{\alpha ,{\mathbf{R}_1}} \sum\limits_{m = 1}^M {{{\left| {\alpha {{\tilde P}_d}\left( {{\theta _m}} \right) - {{\mathbf{a}}^H}\left( {{\theta _m}} \right){\mathbf{Ra}}\left( {{\theta _m}} \right)} \right|}^2}}  \hfill \\
  \displaystyle s.t.\;\;\;\operatorname{diag} \left( {\mathbf{R}}_1 \right) =  \frac{{{P_R}{\mathbf{1}}}}{N}, \hfill \\
  \;\;\;\;\;\;\;\;\;{\mathbf{R}}_1 \succeq 0,{\mathbf{R}}_1 = {{\mathbf{R}}_1^H}, \hfill \\
  \;\;\;\;\;\;\;\;\;\alpha  \ge 0, \hfill \\
  \;\;\;\;\;\;\;\;\;\operatorname{tr} \left( {{\mathbf{f}}_i^*{\mathbf{f}}_i^T{\mathbf{R}}_1} \right) = 0,\forall i,
\end{array}
\end{equation}

\begin{equation}
\begin{gathered}
  \mathop {\min }\limits_{t,{{\mathbf{R}}_1}} \; - t \hfill \\
  s.t.\;{\mathbf{a}}_1^H\left( {{\theta _0}} \right){{\mathbf{R}}_1}{{\mathbf{a}}_1}\left( {{\theta _0}} \right) - {\mathbf{a}}_1^H\left( {{\theta _m}} \right){{\mathbf{R}}_1}{{\mathbf{a}}_1}\left( {{\theta _m}} \right) \ge t,\forall {\theta _m} \in \Omega , \hfill \\
  \;\;\;\;\;\;{\mathbf{a}}_1^H\left( {{\theta _1}} \right){{\mathbf{R}}_1}{{\mathbf{a}}_1}\left( {{\theta _1}} \right) = {\mathbf{a}}_1^H\left( {{\theta _0}} \right){{\mathbf{R}}_1}{{\mathbf{a}}_1}\left( {{\theta _0}} \right)/2, \hfill \\
  \;\;\;\;\;\;{\mathbf{a}}_1^H\left( {{\theta _2}} \right){{\mathbf{R}}_1}{{\mathbf{a}}_1}\left( {{\theta _2}} \right) = {\mathbf{a}}_1^H\left( {{\theta _0}} \right){{\mathbf{R}}_1}{{\mathbf{a}}_1}\left( {{\theta _0}} \right)/2,\; \hfill \\
  \;\;\;\;\;\;{{\mathbf{R}}_1} \succeq 0,{{\mathbf{R}}_1} = {\mathbf{R}}_1^H, \hfill \\
  \;\;\;\;\;\;\operatorname{diag} \left( {{{\mathbf{R}}_1}} \right) = \frac{{{P_R}{\mathbf{1}}}}{{{N}}}, \hfill \\
  \;\;\;\;\;\;\operatorname{tr} \left( {{\mathbf{f}}_i^*{\mathbf{f}}_i^T{{\mathbf{R}}_1}} \right) = 0,\forall i, \hfill \\
\end{gathered}
\end{equation}
respectively, where ${\mathbf{a}}\left( \theta_m  \right) = \left[ {{{\mathbf{a}}_1}\left( \theta_m  \right);{{\mathbf{a}}_2}\left( \theta_m  \right)} \right],\forall m$, and ${{\mathbf{a}}_1}\left( \theta_m  \right)\in {\mathbb{C}^{N_R \times 1}},{{\mathbf{a}}_2}\left( \theta_m  \right) \in {\mathbb{C}^{N_C \times 1}},\forall m$, $P_R$ is the transmission power for radar.
\\\indent Since we assume that the transmitted signals for radar and communication are statistically independent, by recalling (4), it can be easily proven that the overall covariance matrix can be written as
\begin{equation}
{\mathbf{\tilde C}} = \left[ {\begin{array}{*{20}{c}}
  {\mathbf{R}}_1&{\mathbf{0}} \\
  {\mathbf{0}}&{\mathbf{C}}_1
\end{array}} \right] = \left[ {\begin{array}{*{20}{c}}
  {\mathbf{R}}_1&{\mathbf{0}} \\
  {\mathbf{0}}&{{\sum\limits_{k = 1}^K {{{\mathbf{W}}_k}} }}
\end{array}} \right].
\end{equation}
The beampattern gain at $\theta_m$ can be obtained as
\begin{equation}
\begin{gathered}
  {P_d}\left( \theta_m  \right) =  {{\mathbf{a}}^H}\left( \theta_m  \right){\mathbf{\tilde Ca}}\left( \theta_m  \right) \hfill \\
  \;\;\;\;\;\;\;\;\;\;\;\;\;={\mathbf{a}}_1^H\left( \theta_m  \right){\mathbf{R}}_1{{\mathbf{a}}_1}\left( \theta_m  \right) + {\mathbf{a}}_2^H\left( \theta_m \right){\sum\limits_{k = 1}^K {{{\mathbf{W}}_k}} }{{\mathbf{a}}_2}\left( \theta_m  \right). \hfill \\
\end{gathered}
\end{equation}
\\\indent If the shape of the overall beampattern perfectly matches the radar-only beampattern obtained by (12) and (13), we have
\begin{equation}
{\mathbf{a}}_2^H\left( {{\theta _m}} \right)\sum\limits_{k = 1}^K {{{\mathbf{W}}_k}} {{\mathbf{a}}_2}\left( {{\theta _m}} \right) = \sigma {\mathbf{a}}_1^H\left( {{\theta _m}} \right){{\mathbf{R}}_1}{{\mathbf{a}}_1}\left( {{\theta _m}} \right), \forall m,
\end{equation}
where $\sigma\ge 0$ is a scaling factor. Hence, by introducing the notations
\begin{equation}
\begin{gathered}
  {\mathbf{A}} = \left[ {{\mathbf{a}}\left( {{\theta _1}} \right),...,{\mathbf{a}}\left( {{\theta _M}} \right)} \right] \in {\mathbb{C}^{N \times M}}, \hfill \\
  {{\mathbf{A}}_1} = \left[ {{{\mathbf{a}}_1}\left( {{\theta _1}} \right),...,{{\mathbf{a}}_1}\left( {{\theta _M}} \right)} \right] \in {\mathbb{C}^{{N_R} \times M}}, \hfill \\
  {{\mathbf{A}}_2} = \left[ {{{\mathbf{a}}_2}\left( {{\theta _1}} \right),...,{{\mathbf{a}}_2}\left( {{\theta _M}} \right)} \right] \in {\mathbb{C}^{{N_C} \times M}}, \hfill \\
\end{gathered}
\end{equation}
(16) becomes equivalent to
\begin{equation}
\operatorname{diag} \left( {{\mathbf{A}}_2^H\sum\limits_{i = 1}^K {{{\mathbf{W}}_i}} {{\mathbf{A}}_2}} \right) = \sigma \operatorname{diag} \left( {{\mathbf{A}}_1^H{{\mathbf{R}}_1}{{\mathbf{A}}_1}} \right),
\end{equation}
and the downlink beamforming optimization can thus be formulated as
\begin{equation}
\begin{gathered}
  \mathop {\min }\limits_{\sigma ,{{\mathbf{W}}_i}} \;{\left\| {\operatorname{diag} \left( {{\mathbf{A}}_2^H\sum\limits_{i = 1}^K {{{\mathbf{W}}_i}} {{\mathbf{A}}_2} - \sigma {\mathbf{A}}_1^H{{\mathbf{R}}_1}{{\mathbf{A}}_1}} \right)} \right\|^2} \hfill \\
  s.t.\;\;\;{\beta _i} \ge {\Gamma _i},\forall i, \hfill \\
  \;\;\;\;\;\;\;\;{P_1} \le {P_C}, \hfill \\
  \;\;\;\;\;\;\;\;\sigma  \ge 0, \hfill \\
  \;\;\;\;\;\;\;\;{{\mathbf{W}}_i} \succeq 0,{{\mathbf{W}}_i} = {\mathbf{W}}_i^H, \hfill \\
  \;\;\;\;\;\;\;\;\operatorname{rank} \left( {{{\mathbf{W}}_i}} \right) = 1,\forall i, \hfill \\
\end{gathered}
\end{equation}
where $\Gamma_i$ is the SINR threshold of the \emph{i}-th user, $P_1$ and $\beta_i$ are defined by (2) and (3) while $P_C$ is the power budget for downlink communication. It is clear that (19) is non-convex. Nonetheless, a suboptimal solution can be obtained by the classic SDR technique. By omitting the rank-1 constraints, (19) becomes a standard semidefinite program (SDP), which can be efficiently solved. For non-rank-1 solutions, an approximated solution is obtained by standard rank-1 approximation techniques, such as eigenvalue decomposition or Gaussian randomization \cite{5447068}.
\\\indent The proposed zero-forcing optimization can be summarized by the following Algorithm 1.
\renewcommand{\algorithmicrequire}{\textbf{Input:}}
\renewcommand{\algorithmicensure}{\textbf{Output:}}
\begin{algorithm}
\caption{Zero-forcing Beamforming for Separated Deployment}
\label{alg:A}
\begin{algorithmic}
    \REQUIRE $\mathbf{H},\pmb{\Gamma},P_0$, beampattern requirement;
    \ENSURE ${\mathbf{W}}_i,\forall i$;
    \STATE 1. Solve (12) or (13) to obtain the radar covariance matrix ${\bf R}_1\in {\mathbb{C}^{N_R \times N_R}}$;
    \STATE 2. Substitute ${\bf R}_1$ into (19), solve the SDP problem by omitting the rank-1 constraints;
    \STATE 3. Obtain the approximated solution by eigenvalue decomposition or Gaussian randomization.
\end{algorithmic}
\end{algorithm}

\subsection{Beamforming for Shared Deployment}
Although the separated deployment allows flexibility in the design of the radar signal, the need for cancelling its interference inflicted upon the downlink users imposes extra constraints in the radar beampattern design problems, which may result in poor performance. One may also trade-off the radar interference received by users by changing the strict zero-forcing equality constraints to inequalities. However, there are still extra constraints in such problems.
\\\indent By using the shared deployment, the radar's targets of interest can be viewed as virtual downlink users located in a line-of-sight (LoS) channel. The beamforming design thus becomes a power sharing problem between the virtual users in the LoS channel and the real users in the fading channel. The difference is that we meet the requirements of the former by using a specific beampattern, and that of the latter by enforcing their SINR constraints. Therefore, the optimization for the shared deployment is to firstly formulate a radar beampattern by solving (9) or (10) with $N_t = N$, then substitute the covariance matrix obtained in the communication beamforming problem, which is
\begin{subequations}
\begin{align}
  &\mathop {\min }\limits_{{{\mathbf{T}}_i}} \;\left\| {\sum\limits_{i = 1}^K {{{\mathbf{T}}_i}}  - {{\mathbf{R}}_2}} \right\|_F^2 \hfill \\
  &s.t.\;\;\gamma _i \ge {\Gamma _i},\forall i,\;\;\;\;\; \hfill \\
  &\;\;\;\;\;\;\;\operatorname{diag} \left( {\sum\limits_{i = 1}^K {{{\mathbf{T}}_i}} } \right) = \frac{{{P_0}{\mathbf{1}}}}{N},\;\;\;\;\;\; \hfill \\
  &\;\;\;\;\;\;\;{{\mathbf{T}}_i} \succeq 0,{{\mathbf{T}}_i} = {\mathbf{T}}_i^H,\operatorname{rank} \left( {{{\mathbf{T}}_i}} \right) = 1,\forall i,
\end{align}
\end{subequations}
where ${\mathbf{R}}_2$ is obtained by solving (9) or (10), ${P_2}$ and ${\gamma _i}$ are defined as (6) and (7) respectively. Note that the SINR $\gamma_i$ for the \emph{i}-th user in this case is different from $\beta_i$ for the separated deployment above, and that in contrast to (19), all the transmit power is exploited in the above optimization. Here we also employ the equality constraint (20c) for the power budget, since the radar is often required to transmit at the maximum available power in practice\cite{4276989}.
\\\indent For problem (20), the following observations can be made:
\\\indent \emph{Observation 1}: Similar to (19), the problem (20) can be relaxed as SDP by omitting the rank-1 constraints. However, it is difficult to employ eigenvalue decomposition or Gaussian randomization to yield a solution that satisfies the \emph{strict} equality of per-antenna power constraint. Hence, the relaxed solution cannot take full advantage of the available power budget.
\\\indent \emph{Observation 2}: When the number of users becomes large, the DoFs may be insufficient to satisfy the specific constraints. Our results show that for a \emph{K}-user downlink with $K\in \left[N-2, N\right]$, (20) becomes infeasible with a high probability.
\\\indent To avoid these drawbacks, we attempt to simplify (20) as weighted optimizations in conjunction with penalty terms, which can be solved by manifold algorithms.

\subsection{SINR Penalty Terms}
In order to simplify the problem in (20) and to improve the feasibility probability for beamforming optimization, we incorporate the SINR constraints in the objective function as a penalty term. Note that a similar approach can also be developed for the zero-forcing optimization of (19). For notational simplicity, we use $P$ and ${\bf R}$ instead of $P_2$ and ${\bf R}_2$ in the remainder of this paper, unless otherwise specified.
\\\indent 1) \emph{Sum-Square Penalty}: A simple penalty term is the summation of the squared error between the actual values and the thresholds. Therefore, the SINR penalty term can be written as
\begin{equation}
\tilde{\lambda} \left( {\pmb{\gamma }} \right) = \sum\limits_{i = 1}^K {{{\left( {{\gamma _i} - {\Gamma _i}} \right)}^2}}  = {\left\| {{\pmb{\gamma }} - {\pmb{\Gamma }}} \right\|^2},
\end{equation}
where $\pmb\gamma  = {\left[ {{\gamma _1},{\gamma _2},...,{\gamma _K}} \right]^T} \in {\mathbb{R}^{K \times 1}}, \pmb\Gamma  = {\left[ {{\Gamma _1},{\Gamma _2},...,{\Gamma _K}} \right]^T} \in {\mathbb{R}^{K \times 1}}$. Upon recalling the definition for $\gamma_i$ in (7), we have
\begin{equation}
{\gamma _i} - {\Gamma _i} = \frac{{\operatorname{tr} \left( {{{\mathbf{B}}_i}{{\mathbf{T}}_i}} \right)}}{{\operatorname{tr} \left( {{{\mathbf{B}}_i}\sum\limits_{\begin{subarray}{l}
  k = 1 \\
  k \ne i
\end{subarray}}^K {{{\mathbf{T}}_k}} } \right) + {N_0}}} - {\Gamma _i},\forall i,
\end{equation}
where ${\mathbf{B}}_i={\mathbf{h}}_i^*{\mathbf{h}}_i^T$, and ${{\mathbf{T}}_i}$ is defined in (6). Based on above, (21) is equivalent to
\begin{equation}
\lambda \left( \pmb\alpha  \right) = \sum\limits_{i = 1}^K {{{\left( {{\alpha _i} - {\Gamma _i}{N_0}} \right)}^2}}  = {\left\| {\pmb\alpha  - {N_0}\pmb\Gamma } \right\|^2},
\end{equation}
where we have ${\pmb \alpha} = \left[ {{\alpha_1},{\alpha_2},...,{\alpha_K}} \right]^T \in \mathbb{R}^{K\times 1}$, and ${\alpha _i}$ is given by
\begin{equation}
{\alpha _i} = \left( {1 + {\Gamma _i}} \right)\operatorname{tr} \left( {{{\mathbf{B}}_i}{{\mathbf{T}}_i}} \right) - {\Gamma _i}\operatorname{tr} \left( {{{\mathbf{B}}_i}\sum\limits_{k=1}^K {{{\mathbf{T}}_k}} } \right).
\end{equation}
\\\indent 2) \emph{Max Penalty}: To ensure fairness between the downlink users, we can also maximize the minimum SINR of the downlink users. This is formulated as
\begin{equation}
\max\limits_{{{\mathbf{T}}_i}} \;\min\limits_i \;{\gamma _i}.
\end{equation}
In order to transform (25) into a minimization problem, note that
\begin{equation}
\operatorname{arg}\max\limits_{{{\mathbf{T}}_i}} \;\min\limits_i \;{\gamma _i} = \operatorname{arg}\mathop {\min }\limits_{{{\mathbf{T}}_i}} \;\mathop {\max }\limits_i \; - {\gamma _i},
\end{equation}
which yields the following penalty term
\begin{equation}
l\left( {\pmb{\alpha }} \right) = \max \left( { - {\alpha _1}, - {\alpha _2},..., - {\alpha_K}} \right).
\end{equation}
By incorporating (23) and (27) as part of the objective function, the beamforming problem is always feasible. Instead of using the classic SDR approach, we propose to directly solve the non-convex problem by employing manifold-based algorithms, which can ensure that the resultant solution satisfies the equality constraints, and are more computationally efficient than the SDR. The related optimization problems are formulated and solved in the following section.

\section{Beamforming Optimization on the Manifold}

\subsection{Preliminaries on Riemannian Manifold Optimization}
First, let us briefly commence by revisiting the essential concepts of the Riemannian manifold \cite{petersen2006riemannian} that will be useful for the derivation of our algorithms. A manifold $\mathcal{S}$ is a set of points endowed with a locally Euclidean structure near each point. More precisely, each point of an \emph{N}-dimensional manifold has a neighbourhood that is \emph{homeomorphic} to the Euclidean space of dimension \emph{N} \cite{petersen2006riemannian}. Given a point $p$ on $\mathcal{S}$, a \emph{tangent vector} at $p$ is defined as the vector that is tangential to any smooth curves on $\mathcal{S}$ through $p$. As can be seen from Fig. 2, the set of all such vectors at $p$ forms the \emph{tangent space}, denoted by $T_p\mathcal{S}$, which is a Euclidean space.
\\\indent If the tangent spaces of a manifold are equipped with a smoothly varying inner product, the manifold is a \emph{Riemannian manifold}. Accordingly, the associated inner procuct is called a \emph{Riemannian metric}. For instance, $\mathbb{R}^N$ is a Riemannian manifold with the normal vector inner product being the Riemannian metric. It leads to the existence of rich geometric structures on the manifold, namely geodesic distances, curvatures, and gradients of functions.
\\\indent Given a smooth function defined on $\mathcal{S}$, denoted by $f\left(p\right)$, the \emph{Riemannian gradient} is defined as the tangent vector belonging to $T_p\mathcal{S}$ that indicates the steepest ascent direction of $f\left(p\right)$. In particular, if $\mathcal{S}$ is a Riemannian submanifold of a Euclidean space, the Riemannian gradient can be computed as the orthogonal projection of the conventional Euclidean gradient $\nabla {\bar f}\left(p\right)$ onto the tangent space, where ${\bar f}$ is a smooth function defined on the Euclidean space with $f$ being its restriction on $\mathcal{S}$. This is given as \cite{absil2009optimization}
\begin{equation}
\operatorname{grad} f\left( p \right) = {\operatorname{Proj} _p}\nabla {\bar f}\left(p\right).
\end{equation}
For simplicity, we will not distinguish between $f$ and $\bar f$ in the rest of this paper.
\\\indent To find the minimum of $f$, a descent direction is computed based on the Riemannian gradient. For the Riemannian steepest descent (RSD) algorithm, the descent direction is chosen as the negative counterpart of the Riemannian gradient. For the Riemannian conjugate gradient (RCG) algorithm, the descent direction is chosen as a nonlinear combination of the Riemannian gradient of the current iteration and the descent direction of the previous iteration. It is important to note that all the descent directions departing from the point $p$ should belong to the corresponding tangent space $T_p\mathcal{M}$.
\\\indent Once a descent direction is given, the update point for the next iteration is obtained by a specific mapping named as \emph{retraction}, which maps a point on $T_p\mathcal{S}$ to $\mathcal{S}$, with a local rigidity condition that preserves the gradients at $p$ \cite{absil2009optimization}, and is given as
\begin{equation}
{\mathcal{R}_p}\left( \xi   \right) = \tilde p \in \mathcal{S},
\end{equation}
where $\xi \in T_p\mathcal{S}$.
\\\indent Based on the above concepts, below we derive the weighted optimizations related to the original problems of Section III. We will firstly relax the per-antenna power constraint into a total power constraint, i.e. to replace the diagonal constraint with the norm constraint, which is easier to handle by mapping the feasible region onto a hypersphere. This is expected to yield a better solution according to our numerical results. Then we deal with the per-antenna power constraint by solving the problem on an oblique manifold.
\subsection{Beamforming under Total Power Constraint}
\begin{figure}
    \centering
    \includegraphics[width=0.8\columnwidth]{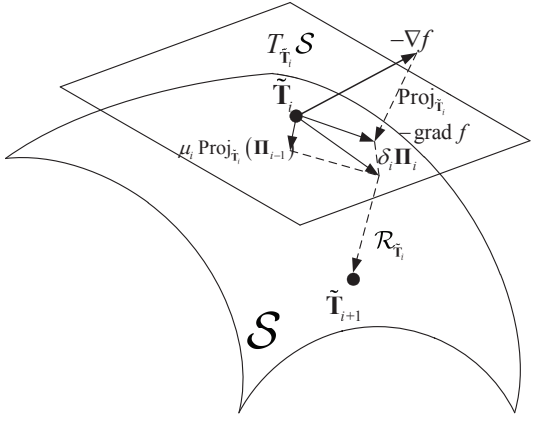}
    \caption{Riemannian conjugate gradient algorithm.}
    \label{fig:2}
\end{figure}

Let us denote the beamforming matrix as ${\mathbf{\tilde T}} = \left[ {{{\mathbf{t}}_1},{{\mathbf{t}}_2},...,{{\mathbf{t}}_K}} \right] \in \mathbb{C}^{N\times K}$. The total power constraint of the problem is thus given by
\begin{equation}
P = \left\| {{\mathbf{\tilde T}}} \right\|_F^2 = \operatorname{tr} \left( {{\mathbf{\tilde T}}{{{\mathbf{\tilde T}}}^H}} \right) = {P_0}.
\end{equation}
Note that the diagonal constraints in both (9) and (10) are also replaced by $\operatorname{tr}\left(\mathbf{R}\right) = P_0$ for this case.
\\\indent 1) \emph{Problem Reformulation}
\\\indent The weighted optimization incorporating the sum-square SINR penalty term is given as
\begin{equation}
\begin{gathered}
  \mathop {\min }\limits_{{\mathbf{\tilde T}}} \;\;{\rho _1}{\left\| {{\mathbf{\tilde T}}{{{\mathbf{\tilde T}}}^H} - {\mathbf{R}}} \right\|^2_F} + {\rho _2}\lambda \left(\pmb{\alpha}  \right) \hfill \\
  s.t.\;\;\;{\left\| {{\mathbf{\tilde T}}} \right\|_F} = \sqrt {{P_0}}, \hfill \\
\end{gathered}
\end{equation}
where $\bf{R}$ is the covariance matrix related to the desired beampattern, $\lambda \left(\pmb{\alpha}  \right)$ is defined by (23), and $\pmb{\rho} = \left[\rho_1,\rho_2 \right]$ is a weighting vector representing the weights for radar and communication in the dual-functional system, respectively. In contrast to the problem (20), there is no rank-1 constraint in the above optimization, where the power constraint is the only constraint. This implies that the feasible set of the problem is a complex hypersphere with the dimension of $NK-1$. Hence, (31) can be reformulated as an unconstrained problem, which is
\begin{equation}
\mathop {\min }\limits_{{\mathbf{\tilde T}} \in \mathcal{S}} \;\;f_1\left(\mathbf{\tilde T} \right)={\rho _1}{\left\| {{\mathbf{\tilde T}}{{{\mathbf{\tilde T}}}^H} - {\mathbf{R}}} \right\|^2_F} + {\rho _2}\lambda \left( \pmb{\alpha}  \right),
\end{equation}
where $\mathcal{S}=\left\{ {{\mathbf{\tilde T}} \in {\mathbb{C}^{N \times K}}\left| {{{\left\| {{\mathbf{\tilde T}}} \right\|}_F}=\sqrt {{P_0}} } \right.} \right\}$ is the complex hypersphere manifold with the radius of $\sqrt {{P_0}}$. Similarly, the max SINR penalty problem can be reformulated as:
\begin{equation}
\mathop {\min }\limits_{{\mathbf{\tilde T}} \in \mathcal{S}} \;\;\tilde{f_2}\left(\mathbf{\tilde T} \right)={\rho _1}{\left\| {{\mathbf{\tilde T}}{{{\mathbf{\tilde T}}}^H} - {\mathbf{R}}} \right\|^2_F} + {\rho _2}l \left( \pmb{\alpha}  \right),
\end{equation}
where $l\left( \pmb{\alpha}  \right)$ is defined by (27). Instead of employing the conventional SDR solver, we propose a Riemannian conjugate gradient algorithm to solve the two non-convex problems by finding the near-global local minima.
\\\indent 2) \emph{Riemannian Gradients}
\\\indent We first compute the Euclidean gradient of $f_1\left(\mathbf{\tilde T} \right)$, which is given by
\begin{equation}
\nabla {f_1}\left( {{\mathbf{\tilde T}}} \right) = \left[ {\frac{{\partial f_1}}{{\partial {{\mathbf{t}}_1}}},\frac{{\partial f_1}}{{\partial {{\mathbf{t}}_2}}},...,\frac{{\partial f_1}}{{\partial {{\mathbf{t}}_K}}}} \right].
\end{equation}
Here we directly give the analytic expression of the gradient, which is
\begin{equation}
\nabla {f_1}\left( {{\mathbf{\tilde T}}} \right) = 4{\rho _1}\left( {{\mathbf{\tilde T}}{{{\mathbf{\tilde T}}}^H} - {\mathbf{R}}} \right){\mathbf{\tilde T}} + 4{\rho _2}\sum\limits_{i = 1}^K {{\alpha _i}{{\mathbf{G}}_i}},
\end{equation}
where $\alpha_i$ is defined by (24), and can be rewritten as
\begin{equation}
{\alpha _i} = \left( {1 + {\Gamma _i}} \right)\operatorname{tr} \left( {{{\mathbf{B}}_i}{{\mathbf{t}}_i}{\mathbf{t}}_i^H} \right) - {\Gamma _i}\operatorname{tr} \left( {{{\mathbf{B}}_i}{\mathbf{\tilde T}}{{{\mathbf{\tilde T}}}^H}} \right),
\end{equation}
while ${\mathbf{G}}_i$ is given by
\begin{equation}
{{\mathbf{G}}_i} = {{\mathbf{B}}_i}\left( {\left( {1 + {\Gamma _i}} \right){{\mathbf{t}}_i}{\mathbf{e}}_i^T - {\Gamma _i}{\mathbf{\tilde T}}} \right).
\end{equation}
In (37), $\mathbf{e}_i \in \mathbb{C}^{K\times1}$ has all-zero entries except for its \emph{i}-th entry, which is equal to 1. For a complex hypersphere, the tangent space is given by \cite{boumal2014manopt}
\begin{equation}
{T_{{\mathbf{\tilde T}}}}\mathcal{S} = \left\{ {{\mathbf{F}} \in {\mathbb{C}^{N \times K}}\left| {\operatorname{Re} \left( {\operatorname{tr} \left( {{{{\mathbf{\tilde T}}}^H}{\mathbf{F}}} \right)} \right) = 0} \right.} \right\}.
\end{equation}
The Riemannian gradient is obtained by projecting the Euclidean gradient onto $T_\mathbf{\tilde T}\mathcal{S}$, which is \cite{boumal2014manopt}
\begin{equation}
\begin{gathered}
  \operatorname{grad} {f_1}\left( {{\mathbf{\tilde T}}} \right) = {\operatorname{Proj} _{{\mathbf{\tilde T}}}}\left( {\nabla {f_1}\left( {{\mathbf{\tilde T}}} \right)} \right) \hfill \\
   = \nabla {f_1}\left( {{\mathbf{\tilde T}}} \right) - \operatorname{Re} \left[ {\operatorname{tr} \left( {{{{\mathbf{\tilde T}}}^H}\nabla {f_1}\left( {{\mathbf{\tilde T}}} \right)} \right)} \right]{\mathbf{\tilde T}}. \hfill \\
\end{gathered}
\end{equation}
\\\indent For the max SINR penalty, unfortunately, the gradient does not exist due to the non-smooth max function. Thus we employ the well-known log-sum-exp upper-bound $\hat{l}\left( \pmb \alpha  \right)$, which is a smooth function, and is given by\cite{7562383}
\begin{equation}
l\left( \pmb \alpha  \right) \le \hat{l}\left( \pmb \alpha  \right) = \varepsilon \log \left( {\sum\limits_{i = 1}^K {\exp \left( {\frac{{ - {\alpha _i}}}{\varepsilon }} \right)} } \right) \le l\left( \pmb \alpha  \right) + \varepsilon \log \left( K \right),
\end{equation}
where $\varepsilon$ is a small positive number.
The approximated max penalty problem is then given as
\begin{equation}
\mathop {\min }\limits_{{\mathbf{\tilde T}} \in \mathcal{S}} \;\;f_2 \left(\mathbf{\tilde T} \right)={\rho _1}{\left\| {{\mathbf{\tilde T}}{{{\mathbf{\tilde T}}}^H} - {\mathbf{R}}} \right\|^2} + {\rho _2}\hat{l} \left( \pmb{\alpha}  \right).
\end{equation}
The Euclidean gradient of $f_2 \left(\mathbf{\tilde T} \right)$ can be computed as
\begin{equation}
\nabla {f_2}\left( {{\mathbf{\tilde T}}} \right) = 4{\rho _1}\left( {{\mathbf{\tilde T}}{{{\mathbf{\tilde T}}}^H} - {\mathbf{R}}} \right){\mathbf{\tilde T}} - 2{\rho _2}\frac{{\sum\limits_{i = 1}^K {\exp \left( { - \frac{{{\alpha _i}}}{\varepsilon }} \right){{\mathbf{G}}_i}} }}{{\sum\limits_{i = 1}^K {\exp \left( { - \frac{{{\alpha _i}}}{\varepsilon }} \right)} }}.
\end{equation}
Accordingly, the Riemannian gradient for $f_2$ is given by
\begin{equation}
\operatorname{grad} {f_2}\left( {{\mathbf{\tilde T}}} \right) = \nabla {f_2}\left( {{\mathbf{\tilde T}}} \right) - \operatorname{Re} \left[ {\operatorname{tr} \left( {{{{\mathbf{\tilde T}}}^H}\nabla {f_2}\left( {{\mathbf{\tilde T}}} \right)} \right)} \right]{\mathbf{\tilde T}}.
\end{equation}
\\\indent 3) \emph{Riemannian Conjugate Gradient Algorithm}
\\\indent Since (32) and (41) are of a similar structure, we can use the RCG to solve both optimizations, where the relevant gradients are given by (35) and (42), respectively. For simplicity, we use $f$ to represent $f_1$ or $f_2$. Let us recall the conventional conjugate gradient (CG) algorithm \cite{wright1999numerical} in the Euclidean space. At step \emph{i+}1, we update the beamforming matrix by \cite{wright1999numerical}
\begin{equation}
{{{\mathbf{\tilde T}}}_{i + 1}} = {{{\mathbf{\tilde T}}}_i} + {\delta _i}{{\pmb \Pi} _i},
\end{equation}
where the stepsize $\delta_i$ is computed by line search algorithms, such as the Armijo rule \cite{wright1999numerical}. The descent direction is obtained as the weighted summation of the current gradient and the descent direction of the last iteration, which is given by
\begin{equation}
{{\pmb \Pi} _i} =  - \nabla f\left( {{{{\mathbf{\tilde T}}}_i}} \right) + {\mu _i}{{\pmb \Pi} _{i - 1}},
\end{equation}
where $\mu_i$ is given by the Polak-Ribi\`ere formula\cite{wright1999numerical}
\begin{equation}
{\mu _i} = \frac{{\left\langle {\nabla f\left( {{{{\mathbf{\tilde T}}}_i}} \right),\nabla f\left( {{{{\mathbf{\tilde T}}}_i}} \right) - \nabla f\left( {{{{\mathbf{\tilde T}}}_{i - 1}}} \right)} \right\rangle }}{{\left\langle {\nabla f\left( {{{{\mathbf{\tilde T}}}_{i - 1}}} \right),\nabla f\left( {{{{\mathbf{\tilde T}}}_{i - 1}}} \right)} \right\rangle}}.
\end{equation}
Here, ${\left\langle \cdot,\cdot \right\rangle}$ denotes the usual Euclidean inner product on $\mathbb{C}^{N\times K}$, which is
\begin{equation}
\left\langle {{\mathbf{X}},{\mathbf{Y}}} \right\rangle = \operatorname{Re} \left[ {\operatorname{tr} \left( {{{\mathbf{X}}^H}{\mathbf{Y}}} \right)} \right].
\end{equation}
For the RCG method, as mentioned previously, the beamforming matrix to be updated at step \emph{i+}1 is obtained by the retraction on the complex hypersphere, which is
\begin{equation}
{{{\mathbf{\tilde T}}}_{i + 1}} = {\mathcal{R}_{{{{\mathbf{\tilde T}}}_i}}}\left( {{\delta _i}{{\mathbf{\Pi }}_i}} \right) = \frac{{\sqrt {{P_0}} \left( {{{{\mathbf{\tilde T}}}_{i}} + {\delta _i}{{\mathbf{\Pi }}_i}} \right)}}{{{{\left\| {{{{\mathbf{\tilde T}}}_{i}} + {\delta _i}{{\mathbf{\Pi }}_i}} \right\|}_F}}}.
\end{equation}
To compute $\mu_i$, we note that for the RCG method, $\operatorname{grad} {f}\left( {{\mathbf{\tilde T}}_i} \right)$ and $\operatorname{grad} {f}\left( {{\mathbf{\tilde T}}_{i-1}} \right)$ belong to the different tangent spaces ${T_{{\mathbf{\tilde T}}_i}}\mathcal{S}$ and ${T_{{\mathbf{\tilde T}}_{i-1}}}\mathcal{S}$, hence they cannot be simply added as in (46). Following \cite{boumal2014thesis}, we use \emph{vector transport} to perform the nonlinear combination, by mapping a vector from a tangent space to the vector on another tangent space. For the complex hypersphere, we simply define the projector used in (39) as the vector transport. Hence, to transport  $\operatorname{grad} {f}\left( {{\mathbf{\tilde T}}_{n-1}} \right)\in {T_{{\mathbf{\tilde T}}_{n-1}}}\mathcal{S}$ to ${T_{{\mathbf{\tilde T}}_n}}\mathcal{S}$ is to project it onto  ${T_{{\mathbf{\tilde T}}_n}}\mathcal{S}$, which is
\begin{equation}
{\mathcal{T}_{{{{\mathbf{\tilde T}}}_{i - 1}}}}\left[ {\operatorname{grad} f\left( {{{{\mathbf{\tilde T}}}_{i - 1}}} \right)} \right] = {\operatorname{Proj} _{{{{\mathbf{\tilde T}}}_i}}}\left[ {\operatorname{grad} f\left( {{{{\mathbf{\tilde T}}}_{i - 1}}} \right)} \right].
\end{equation}
Then it follows that
\begin{equation}
\begin{small}
\begin{gathered}
  {\mu _i} \hfill \\
   = \frac{{{{\left\langle {\operatorname{grad} f\left( {{{{\mathbf{\tilde T}}}_i}} \right),\operatorname{grad} f\left( {{{{\mathbf{\tilde T}}}_i}} \right) - {\mathcal{T}_{{{{\mathbf{\tilde T}}}_{i - 1}}}}\left( {\operatorname{grad} f\left( {{{{\mathbf{\tilde T}}}_{i - 1}}} \right)} \right)} \right\rangle }_R}}}{{{{\left\langle {\operatorname{grad} f\left( {{{{\mathbf{\tilde T}}}_{i - 1}}} \right),\operatorname{grad} f\left( {{{{\mathbf{\tilde T}}}_{i - 1}}} \right)} \right\rangle }_R}}}, \hfill \\
\end{gathered}
\end{small}
\end{equation}
where ${\left\langle \cdot,\cdot \right\rangle}_R$ is the Riemannian metric, which is chosen as the metric induced from the Euclidean space, i.e., the same inner product defined in (47). Similarly, ${\mathbf{\Pi }}_i$ is obtained as
\begin{equation}
{\pmb{\Pi} _i} =  - \operatorname{grad} f\left( {{{{\mathbf{\tilde T}}}_i}} \right) + {\mu _i}{\mathcal{T}_{{{{\mathbf{\tilde T}}}_{i-1}}}}\left( {{\pmb{\Pi} _{i - 1}}} \right).
\end{equation}
In Fig. 2, we show the procedure of the RCG algorithm for a single iteration, which is also summarized by Algorithm 2.
\\\indent \emph{Remark}: In contrast to the SDR method, which searches the solution in the positive semidefinite cone and performs a rank-1 approximation afterwards, the proposed Algorithm 2 searches directly on the hypersphere without approximation, and therefore it is more efficient. Additionally, if the solution obtained by the SDP has a rank higher than 1, the approximated rank-1 solution may not satisfy the strict equality power constraint. By contrast, the proposed RCG algorithm guarantees to satisfy the power constraint.
\renewcommand{\algorithmicrequire}{\textbf{Input:}}
\renewcommand{\algorithmicensure}{\textbf{Output:}}
\begin{algorithm}
\caption{RCG Algorithm for Solving (32) and (41)}
\label{alg:B}
\begin{algorithmic}
    \REQUIRE $\mathbf{R}, \mathbf{H},\pmb{\Gamma}, \pmb{\rho}=\left[\rho_1,\rho_2\right],P_0, \Delta >0, i_{max}>0$
    \ENSURE $\tilde{\mathbf{T}}_i$
    \STATE 1. Initialize randomly $\tilde{\mathbf{T}}_0 \in \mathcal{S}$, \\ set $\mathbf{\Pi}_0=-\operatorname{grad}f\left(\tilde{\mathbf{T}}_0\right), i=0$.
    \WHILE{$i\le i_{max}$ and ${\left\| {\operatorname{grad} f\left( {{{{\mathbf{\tilde T}}}_{i}}} \right)} \right\|_F} \ge {\Delta}$}
    \STATE 2. $i=i+1$.
    \STATE 3. Compute stepsize $\delta_{i-1}$ by Armijo rule, and set ${{{\mathbf{\tilde T}}}_{i}}$ by the retraction (48).
    \STATE 4. Compute $\mu_i$ by (50).
    \STATE 5. Compute ${\pmb{\Pi} _i}$ by (51).
    \ENDWHILE
\end{algorithmic}
\end{algorithm}

\subsection{Beamforming under Per-Antenna Power Constraint}
We now use the RCG algorithm to solve problems under per-antenna power constraints. Upon letting $\mathbf{X} = {\mathbf{\tilde T}}^H$, it follows that
\begin{equation}
\begin{gathered}
  \left\| {{\mathbf{\tilde T}}{{{\mathbf{\tilde T}}}^H} - {\mathbf{R}}} \right\|_F^2 = \left\| {{{\mathbf{X}}^H}{\mathbf{X}} - {\mathbf{R}}} \right\|_F^2, \hfill \\
  \operatorname{diag} \left( {{\mathbf{\tilde T}}{{{\mathbf{\tilde T}}}^H}} \right) = \operatorname{diag} \left( {{{\mathbf{X}}^H}{\mathbf{X}}} \right). \hfill \\
\end{gathered}
\end{equation}
Note that here the original covariance matrix $\mathbf{R}$ is obtained by solving the MIMO radar beamforming problem of (9) or (10) under diagonal constraints. The sum-square SINR penalty problem is then reformulated as
\begin{equation}
\begin{gathered}
  \mathop {\min}\limits_{\mathbf{X}} \;{\rho _1}\left\| {{{\mathbf{X}}^H}{\mathbf{X}} - {\mathbf{R}}} \right\|_F^2 + {\rho _2}\lambda \left( {\pmb{\alpha }} \right) \hfill \\
  s.t.\;\;\operatorname{diag} \left( {{{\mathbf{X}}^H}{\mathbf{X}}} \right) = \frac{{{P_0}{\mathbf{1}}}}{N}. \hfill \\
\end{gathered}
\end{equation}
In this case, $\alpha_i$ can be rewritten as
\begin{equation}
{\alpha _i} = \left( {1 + {\Gamma _i}} \right)\operatorname{tr} \left[ {{{\mathbf{B}}_i}{{\mathbf{X}}^H}\left( {i,:} \right){\mathbf{X}}\left( {i,:} \right)} \right] - {\Gamma _i}\operatorname{tr} \left( {{{\mathbf{B}}_i}{{\mathbf{X}}^H}{\mathbf{X}}} \right),
\end{equation}
where ${\mathbf{X}}\left( {i,:} \right)$ denotes the \emph{i}-th row of ${\mathbf{X}}$.
The feasible region of (53) can be viewed as a complex oblique manifold, which is formulated as
\begin{equation}
\mathcal{M} = \left\{ {{\mathbf{X}} \in {\mathbb{C}^{K \times N}}\left| {{{({{\mathbf{X}}^H}{\mathbf{X}})}_{nn}} = \frac{{{P_0}}}{N},\forall n} \right.} \right\},
\end{equation}
where ${{({{\mathbf{X}}^H}{\mathbf{X}})}_{nn}}$ is the \emph{n}-th diagonal entry of ${{({{\mathbf{X}}^H}{\mathbf{X}})}}$. By the above definition, the penalty problems can be equivalently written as
\begin{equation}
\mathop {\min }\limits_{{\mathbf{X}} \in \mathcal{M}} \;{f_3}\left( {\mathbf{X}} \right) = {\rho _1}\left\| {{{\mathbf{X}}^H}{\mathbf{X}} - {\mathbf{R}}} \right\|_F^2 + {\rho _2}\lambda \left( {\pmb{\alpha }} \right),
\end{equation}
\begin{equation}
\mathop {\min }\limits_{{\mathbf{X}} \in \mathcal{M}} \;{f_4}\left( {\mathbf{X}} \right) = {\rho _1}\left\| {{{\mathbf{X}}^H}{\mathbf{X}} - {\mathbf{R}}} \right\|_F^2 + {\rho _2}\hat{l} \left( {\pmb{\alpha }} \right).
\end{equation}
Therefore, (56) and (57) can also be regarded as unconstrained problems defined on a Riemannian manifold $\mathcal{M}$. Similar to the case of total power constraints, we first compute the Euclidean gradient for ${f_3}\left( {\mathbf{X}} \right)$ and ${f_4}\left( {\mathbf{X}} \right)$ as
\begin{equation}
\nabla {f_3}\left( {\mathbf{X}} \right) = 4{\rho _1}{\mathbf{X}}\left( {{{\mathbf{X}}^H}{\mathbf{X}} - {\mathbf{R}}} \right) + 4{\rho _2}\sum\limits_{i = 1}^K {{\alpha _i}{\mathbf{G}}_i^H},
\end{equation}
\begin{equation}
\nabla {f_4}\left( {\mathbf{X}} \right) = 4{\rho _1}{\mathbf{X}}\left( {{{\mathbf{X}}^H}{\mathbf{X}} - {\mathbf{R}}} \right) - 2{\rho _2}\frac{{\sum\limits_{i = 1}^K {\exp \left( { - \frac{{{\alpha _i}}}{\varepsilon }} \right){\mathbf{G}}_i^H} }}{{\sum\limits_{i = 1}^K {\exp \left( { - \frac{{{\alpha _i}}}{\varepsilon }} \right)} }},
\end{equation}
where ${\mathbf{G}}_i$ is defined by (37) and can be simply transformed as a function of $\mathbf{X}$.
\\\indent Accordingly, the Riemannian gradients corresponding to the Euclidean gradients (58) and (59) are given by projection on the tangent space of $\mathcal{M}$, which is\cite{boumal2014manopt}
\begin{equation}
{T_{\mathbf{X}}}\mathcal{M} = \left\{ {{\mathbf{F}} \in {\mathbb{C}^{K \times N}}\left| {\operatorname{Re} \left( {{{\left( {{{\mathbf{X}}^H}{\mathbf{F}}} \right)}_{nn}}} \right) = 0,\forall n} \right.} \right\}.
\end{equation}
The Riemannian gradients are thus given as\cite{boumal2014manopt}
\begin{equation}
\begin{gathered}
  \operatorname{grad} {f_3}\left( {\mathbf{X}} \right) = {\operatorname{Proj} _{\mathbf{X}}}\nabla {f_3}\left( {\mathbf{X}} \right) \hfill \\
  \;\;\;\;\;\;\;\;\;\;\;\;\;\;\;\;\;\;\; = \nabla {f_3}\left( {\mathbf{X}} \right) - {\mathbf{X}}\operatorname{ddiag} \left[ {\operatorname{Re} \left( {{{\mathbf{X}}^H}\nabla {f_3}\left( {\mathbf{X}} \right)} \right)} \right], \hfill \\
   \operatorname{grad} {f_4}\left( {\mathbf{X}} \right) = {\operatorname{Proj} _{\mathbf{X}}}\nabla {f_4}\left( {\mathbf{X}} \right) \hfill \\
  \;\;\;\;\;\;\;\;\;\;\;\;\;\;\;\;\;\;\; = \nabla {f_4}\left( {\mathbf{X}} \right) - {\mathbf{X}}\operatorname{ddiag} \left[ {\operatorname{Re} \left( {{{\mathbf{X}}^H}\nabla {f_4}\left( {\mathbf{X}} \right)} \right)} \right]. \hfill \\
\end{gathered}
\end{equation}
Following a similar procedure to that of Algorithm 2, we calculate the (\emph{i}+1)-st update by the following retraction
\begin{equation}
\begin{gathered}
  {{\mathbf{X}}_{i + 1}} = {\mathcal{R}_{{{\mathbf{X}}_i}}}\left( {{\eta _i}{{\mathbf{J}}_i}} \right) \hfill \\
  \;\;\;\;\;\;\;\;\; = \sqrt{\frac{P_0}{N}}\left[ {\frac{{{{\left( {{{\mathbf{X}}_i} + {\eta _i}{{\mathbf{J}}_i}} \right)}_1}}}{{\left\| {{{\left( {{{\mathbf{X}}_i} + {\eta _i}{{\mathbf{J}}_i}} \right)}_1}} \right\|}},...,\frac{{{{\left( {{{\mathbf{X}}_i} + {\eta _i}{{\mathbf{J}}_i}} \right)}_N}}}{{\left\| {{{\left( {{{\mathbf{X}}_i} + {\eta _i}{{\mathbf{J}}_i}} \right)}_N}} \right\|}}} \right],
\end{gathered}
\end{equation}
where the stepsize $\eta_i$ is obtained by Armijo rule, the descent direction $\mathbf{J}_i$ is computed by
\begin{equation}
{\mathbf{J}_i} =  - \operatorname{grad} f\left( {{{{\mathbf{X}}}_i}} \right) + {\tau _i}{\mathcal{T}_{{{{\mathbf{X}}}_{i-1}}}}\left( {{\mathbf{J} _{i - 1}}} \right),
\end{equation}
where $f$ represents both $f_3$ and $f_4$. By defining the vector transport on $\mathcal{M}$ as
\begin{equation}
{\mathcal{T}_{{{\mathbf{X}}_{i - 1}}}}\left[ {\operatorname{grad} f\left( {{{\mathbf{X}}_{i - 1}}} \right)} \right] = {\operatorname{Proj} _{{{\mathbf{X}}_i}}}\left[ {\operatorname{grad} f\left( {{{\mathbf{X}}_{i - 1}}} \right)} \right],
\end{equation}
$\tau _i$ can be obtained by
\begin{equation}
\begin{small}
\begin{gathered}
  {\tau _i} \hfill \\
   = \frac{{{{\left\langle {\operatorname{grad} f\left( {{{{\mathbf{X}}}_i}} \right),\operatorname{grad} f\left( {{{{\mathbf{X}}}_i}} \right) - {\mathcal{T}_{{{{\mathbf{X}}}_{i - 1}}}}\left[ {\operatorname{grad} f\left( {{{{\mathbf{X}}}_{i - 1}}} \right)} \right]} \right\rangle }_R}}}{{{{\left\langle {\operatorname{grad} f\left( {{{{\mathbf{X}}}_{i - 1}}} \right),\operatorname{grad} f\left( {{{{\mathbf{X}}}_{i - 1}}} \right)} \right\rangle }_R}}}. \hfill \\
\end{gathered}
\end{small}
\end{equation}
We summarize the above procedure in Algorithm 3.

\renewcommand{\algorithmicrequire}{\textbf{Input:}}
\renewcommand{\algorithmicensure}{\textbf{Output:}}
\begin{algorithm}
\caption{RCG Algorithm for Solving (56) and (57)}
\label{alg:C}
\begin{algorithmic}
    \REQUIRE $\mathbf{R}, \mathbf{H},\pmb{\Gamma}, \pmb{\rho}=\left[\rho_1,\rho_2\right],P_0, \Delta >0, i_{max}>0$
    \ENSURE $\mathbf{X}_i$
    \STATE 1. Initialize randomly $\mathbf{X}_0 \in \mathcal{M}$, \\ set $\mathbf{J}_0=-\operatorname{grad}f\left(\mathbf{X}_0\right), i=0$.
    \WHILE{$i\le i_{max}$ and ${\left\| {\operatorname{grad} f\left( {{{{\mathbf{X}}}_{i}}} \right)} \right\|_F} \ge {\Delta}$}
    \STATE 2. $i=i+1$.
    \STATE 3. Compute stepsize $\eta_{i-1}$ by Armijo rule, and set ${{{\mathbf{X}}}_{i}}$ by the retraction (62).
    \STATE 4. Compute $\tau_i$ by (65).
    \STATE 5. Compute ${\mathbf{J} _i}$ by (63).
    \ENDWHILE
\end{algorithmic}
\end{algorithm}

\section{Complexity Analysis}
In this section, we study the computational complexity of using the RCG algorithm to solve the proposed optimization problems in terms of flops. We note that since there are no closed-form complexity expressions for the SDR algorithms, we cannot analytically compare the two families. Hence a simulation based comparison is presented in the numerical results, which highlights the complexity savings of the RCG algorithms.
\\\indent Following the concept of\cite{boyd2004convex}, a flop is defined as one addition, subtraction, multiplication, or division of two floating-point numbers. With this definition, one complex addition and one complex multiplication can be viewed as 2 and 6 flops, respectively. One addition between two $M \times N$ complex matrices includes $2MN$ flops, and one multiplication between a $M \times N$ and a $N \times P$ complex matrix needs $8MNP$ flops. Since the number of iterations is difficult to predict, we consider the complexity per iteration, and study the convergence of the algorithms in terms of iterations numerically in our results. In all the computations, we only consider the highest-order terms and omit the low-order ones. For clarity, we list the complexity per iteration for both algorithms in TABLE I by each operation.

\begin{table}
\renewcommand{\arraystretch}{1.3}
\caption{Computational Complexity per Iteration for Algorithm 2 and 3}
\label{table_example}
\centering
\begin{tabular}{lr}
\toprule
 \bf Operation & \bf Flops \\
\midrule
Retraction & $14NK$  \\
Euclidean Gradient & $23N^2K+12NK^2$  \\
Riemannian Gradient & $12NK$ \\
Vector Transport & $12NK$   \\
Inner Product & $8NK$  \\
\hline
\bf Total &  $23N^2K+12NK^2$  \\
\bottomrule
\end{tabular}
\end{table}

\subsection{Complexity of Beamforming Problems under Total Power Constraint}
We first consider the complexity of solving (32). For the retraction of line 3 in Algorithm 2, which is computed by (48), the total number of flops needed is $14NK$. For line 4, the costs of computing the Euclidean and Riemannian gradient using (35) and (39) are $23N^2K+12NK^2$ and $12NK$ flops. The vector transport (49) and the inner product (47) need $12NK$ and $8NK$ flops, respectively. Hence, the costs for line 4 are $23N^2K+12NK^2+42NK$ flops. For line 5, the costs are $16NK$ flops, which lead to a total of $23N^2K+12NK^2$ flops for each iteration.
\\\indent For problem (41), we employ the log-sum-exp approximation, and therefore the only difference w.r.t. (32) is to compute the Euclidean gradient. Here we take the exponential of a floating-point number as an operation imposing a constant number of flops, which leads to the same order of complexity as a simple addition or multiplication. Hence, by omitting the lower-order terms, the costs for computing the gradient (42) are the same as those of (35), which means that the total costs for solving (41) are also $23N^2K+12NK^2$ flops per iteration.
\subsection{Complexity of Beamforming Problems under Per-Antenna Power Constraint}
For solving (56) and (57), the costs of retraction (62) are also $14NK$ flops. The costs for Euclidean gradients are the same as those of (35) and (42), respectively. Noting that the operators $\operatorname{ddiag}\left(\mathbf{X}\mathbf{Y}\right)$ and $\operatorname{tr}\left(\mathbf{X}\mathbf{Y}\right)$ involve a complexity of the same order, the costs needed for computing the corresponding Riemannian gradient, the vector transport and the inner product are the same as those of the optimizations under total power constrains. Therefore, the overall costs of solving (56) and (57) are also $23N^2K+12NK^2$ flops in each iteration.
\\\indent It can be seen from the above discussions that the computational complexities for solving both beamforming problems are $\mathcal{O}\left(N^2K+NK^2\right)$ in each iteration.

\section{Numerical Results}
In this section, numerical results based on Monte Carlo simulations are provided for validating the efficiency of the proposed beamforming approaches. In all the simulations, we set $P_0 = 20\text{dBm}, N = 20$, $N_0 = 0\text{dBm}$, and employ a ULA with half-wavelength spacing between adjacent antennas. We also assume that each entry of the channel matrix $\mathbf{H}$ obeys the i.i.d. standard complex Gaussian distribution. We solve the constrained beamforming optimizations of Section III-B and C by the classic SDR technique using the CVX toolbox \cite{grant2008cvx}, and the penalty problems of Section IV-B and C by the RCG algorithm. From Fig. 3 to Fig. 5, we denote the beampatterns obtained for radar and RadCom joint transmission by `Radar-Only' and `RadCom', respectively. In the rest of the figures, where the performance of different beamforming methods is compared for the shared deployment, we use `Constrainted' to denote the constrained optimization (20), `RCG' and `SDR' for the algorithm that is employed, i.e. the Riemannian conjugate gradient algorithm and the semidefinite relaxation, `Total' and `Per-Ant' for the total power constraint and per-antenna power constraint, `Max' and `Sum-Squ' for the max SINR and sum-square SINR penalty terms used in the weighted optimizations.
\subsection{Comparison between the Separated and Shared Deployments}
In order to evaluate the performance of the two proposed antenna deployments, we compare the beampatterns obtained by the zero-forcing beamforming and by the shared beamforming. Fig 3 (a) and (b) show the beampatterns with multi-beams, which are originally obtained by solving the problems of (12) and (9) for the radar-only beamforming, and are then formulated by solving (19) and (20) for the two RadCom cases, respectively. The locations of the 5 beams are $\left[-60^\circ,-36^\circ,0^\circ,36^\circ,60^\circ \right]$. The total transmit power, the required SINR for each user and the number of users are set as $\Gamma = 10\text{dB}, K = 4$. For the separated deployment, we set $N_R = 14, N_C = 6,P_R = P_C = P_0/2$. It can be seen that the separated deployment provides a poor beampattern with low peaks at each beam owing to its lower DoF, while the shared deployment achieves a far better one, with even higher peaks than the radar-only beampattern. In Fig. 4 (a) and (b), we investigate the performance of the two deployments in the case of their 3dB beampattern formulations, which are originally obtained by solving problems (13) and (10) for radar only and are then formulated by solving (19) and (20) for the RadCom cases, where the main-beam is centered at $0^\circ$ with a 3dB width of $10^\circ$. All other parameters remain the same as in Fig. 3. Note that the zero-forcing beamforming formulates a RadCom beampattern with a peak-sidelobe-ratio (PSLR) of 7dB, while the shared beamforming achieves a PSLR of 15dB. We also show in Fig. 5 the trade-off between PSLR and SINR for the resultant 3dB beampattern designs of the two deployments, where other parameters are the same as in Fig. 3 and Fig. 4. Once again, we see that for a fixed SINR level, the shared deployment outperforms the separated case leading to a substantial 8dB gain in PSLR.
\begin{figure}
\centering
\subfloat[]{\includegraphics[width=0.85\columnwidth]{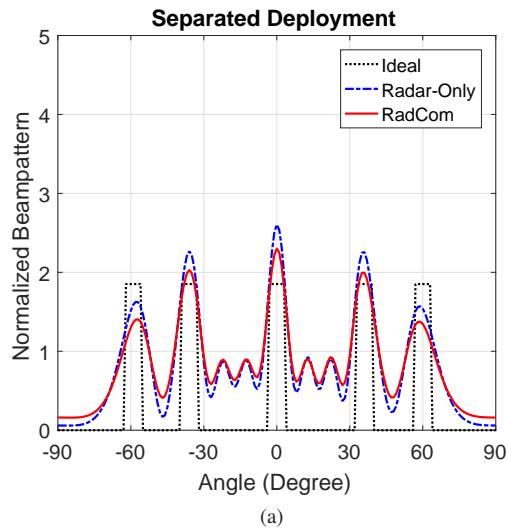}
\label{fig.4}}
\vspace{0.1in}
\hspace{.1in}
\subfloat[]{\includegraphics[width=0.85\columnwidth]{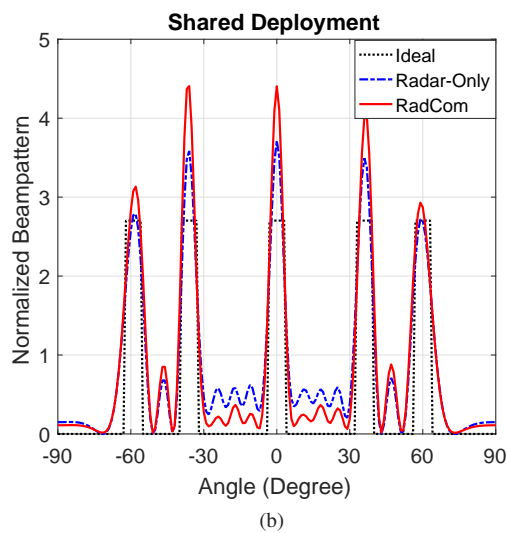}
\label{fig.5}}
\caption{Multi-beam beampatterns comparisons for $\Gamma = 10\text{dB}, K = 4$. (a) Separated deployment; (b) Shared deployment.}
\label{fig_shared_ZF_comparison_MB}
\end{figure}

\begin{figure}
\subfloat[]{\includegraphics[width=0.85\columnwidth]{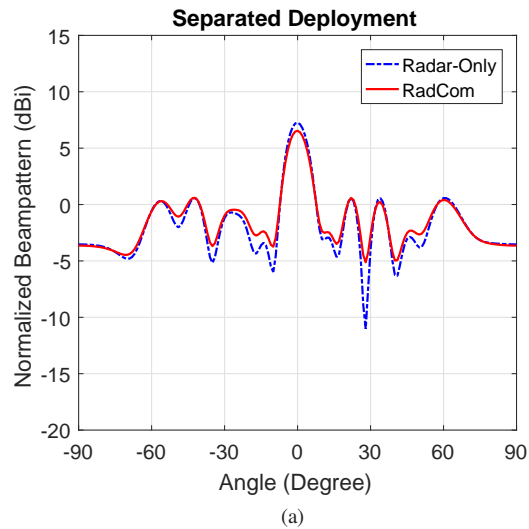}
\label{fig.6}}
\vspace{0.1in}
\hspace{.1in}
\subfloat[]{\includegraphics[width=0.85\columnwidth]{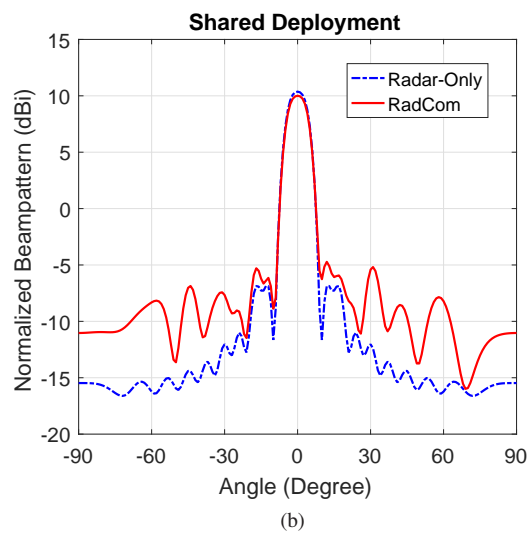}
\label{fig.7}}
\caption{3dB beampatterns comparisons for $\Gamma = 10\text{dB}, K = 4$. (a) Separated deployment; (b) Shared deployment.}
\label{fig_shared_ZF_comparison}
\end{figure}
\begin{figure}
    \centering
    \includegraphics[width=0.85\columnwidth]{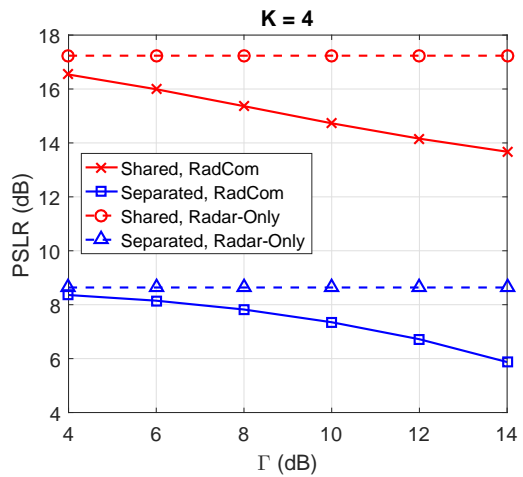}
    \caption{Trade-off between PSLR and SINR level, $P_0 = 20\text{dBm}, K = 4$.}
    \label{fig:8}
\end{figure}

\subsection{Feasibility Comparison between Constrained Problems and Penalty Problems}
In Fig. 6, we show the feasibility probability of the constrained problems and weighted problems with $K$ ranging from 17 to 20, where $\Gamma = 10\text{dB}$. Note that the feasibility probability of the problem (20) decreases with the increase of the number of users due to the reduction of DoFs. For the case that the number of users equals to the number of the BS antennas, the feasibility probability is less than 5\%. For the zero-forcing beamforming of the separated deployment, the optimization is generally infeasible for large numbers of users. Nevertheless, all the weighted optimizations are always feasible, since they are unconstrained problems on the manifolds.
\begin{figure}
    \centering
    \includegraphics[width=0.85\columnwidth]{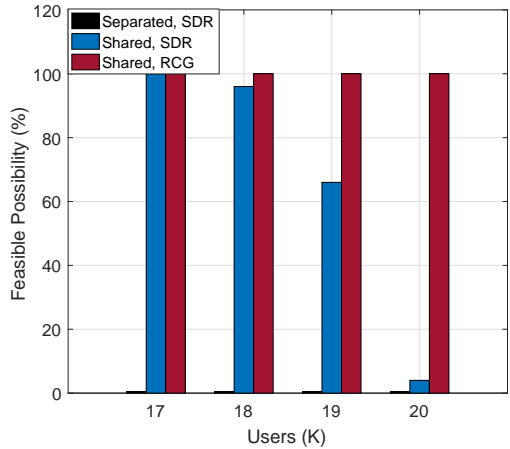}
    \caption{Feasibility probability comparison between constrained problems and penalty problems, $\Gamma = 10\text{dB}$.}
    \label{fig:9}
\end{figure}

\subsection{Performance of Beamforming Methods for Shared Deployment}
\begin{figure}
    \centering
    \includegraphics[width=0.85\columnwidth]{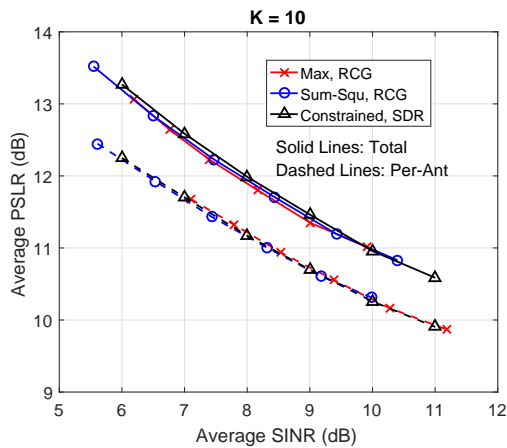}
    \caption{Trade-off between PSLR and SINR for different methods, $K = 10$.}
    \label{fig:10}
\end{figure}
\begin{figure}
    \centering
    \includegraphics[width=0.85\columnwidth]{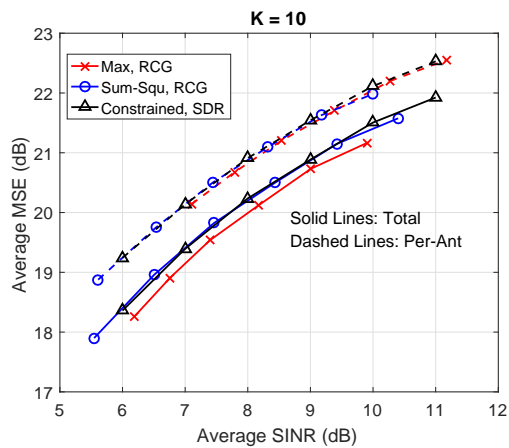}
    \caption{Trade-off between MSE and SINR for different methods, $K = 10$.}
    \label{fig:11}
\end{figure}
Figures 7 and 8 investigate the performance trade-off between radar and communications for different beamforming methods with the shared deployment. In both figures, an original 3dB beampattern of the same shape as in Fig. 4 (b) is generated, while SDR and RCG algorithms are used to solve the corresponding constrained and weighted optimizations. We fix the number of users at $K = 10$, and represent the beamforming approaches under total power and per-antenna power constraints by solid and dashed lines, respectively. The weighting vectors of the different weighted optimizations are given in TABLE II. Fig. 7 presents the trade-off between the PSLR of the RadCom beampattern and the average SINR of the downlink users. Based on this two observations can be made: 1) As expected, the beamforming methods under total power constraints outperform the methods that employ per-antenna power constraints with a 1.7dB SINR gain; 2) The weighted optimizations achieve nearly the same performance as the constrained optimizations, which indicates that the performance loss of penalty problems against the constrained problems is negligible. In Fig. 8, the trade-offs between the mean squared error (MSE) of beampatterns and the average downlink SINR are characterized. Here the squared error is defined as $\sum\limits_{m = 1}^M {{{\left( {{P_{{\text{radar}}}}\left( {{\theta _m}} \right) - {P_{{\text{RadCom}}}}\left( {{\theta _m}} \right)} \right)}^2}}$, where ${{P_{{\text{radar}}}}\left( {{\theta _m}} \right)}$ and ${{P_{{\text{RadCom}}}}\left( {{\theta _m}} \right)}$ denote the beampatterns for radar-only and RadCom cases, respectively. Unsurprisingly, we see the similar trends to those in Fig. 7. It is also worth noting that for beamforming methods under total power constraints, the weighted optimization with the max SINR penalty term performs even better than the constrained optimization.
\begin{table}
\renewcommand{\arraystretch}{1.3}
\caption{Weighting Vectors for Weighted Optimizations of Fig. 7, 8, 11 and 12}
\label{table_2}
\centering
\begin{tabular}{ccc}
\toprule
\bf   & \bf  Total Constraint & \bf Per-Ant Constraint\\
\midrule
 & $\left[\rho_1,\rho_2\right]$ & $\left[\rho_1,\rho_2\right]$ \\
\bf Sum Square Penalty & $\left[10,1\right]$ & $\left[3,1\right]$  \\
\bf Max Penalty & $\left[10,1\right]$ & $\left[1,2\right]$ \\
\bottomrule
\end{tabular}
\end{table}

\begin{table}
\renewcommand{\arraystretch}{1.3}
\caption{Weighting Vectors for Weighted Optimizations of Fig. 9}
\label{table_3}
\centering
\begin{tabular}{ccc}
\toprule
\bf   & \bf Total Constraint & \bf Per-Ant Constraint\\
\midrule
 & $\left[\rho_1,\rho_2\right]$ & $\left[\rho_1,\rho_2\right]$ \\
\bf Sum Square Penalty & $\left[10,1\right]$ & $\left[7,1\right]$  \\
\bf Max Penalty & $\left[58,1\right]$ & $\left[3,1\right]$ \\
\bottomrule
\end{tabular}
\end{table}

We further note that, in contrast to the constrained optimizations, the SINR obtained for weighted optimizations is not explicitly constrained. To evaluate the impact of this, in Fig. 9 we look at the probability distribution of the instantaneous SINR based on the actual transmitted symbols for the different methods in conjunction with $K = 4$ and $\Gamma = 18\text{dB}$. The weighting vectors of TABLE III are employed for achieving the target SINR. It can be observed that the instantaneous SINR ranges from 14dB to 22dB for both constrained and weighted optimizations. This highlights that even under explicit SINR constraints, which impose an average SINR threshold, the instantaneous SINR obtained may indeed vary. Still, it can be seen that although the overall variation in the SINR obtained is similar for both weighted and constrained optimizations, and that the sum-square SINR penalty problem associated with the total power constraint achieves the equivalent variance of the corresponding constrained version, the distributions of the constrained optimizations are in general closer to the mean value than for the weighted optimizations. Nevertheless, we will show in the following that this variation is affordable in the light of the significant reduction of the computational complexity.

\begin{figure}
    \centering
    \includegraphics[width=0.85\columnwidth]{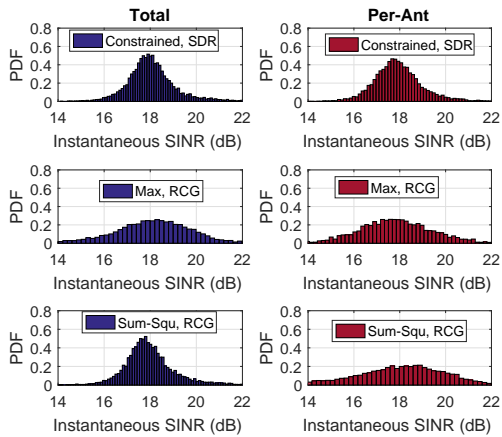}
    \caption{Instantaneous SINR distributions for different methods,  $K = 10, \Gamma = 10\text{dB}$.}
    \label{fig:15}
\end{figure}

\subsection{Computational Complexity}
\begin{figure}
    \centering
    \includegraphics[width=0.85\columnwidth]{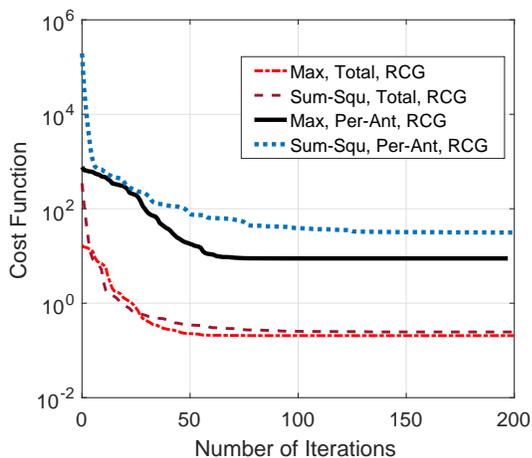}
    \caption{Cost functions vs. number of iterations for different algorithms, $K = 6, \Gamma = 10\text{dB}$.}
    \label{fig:12}
\end{figure}
\begin{figure}
    \centering
    \includegraphics[width=0.85\columnwidth]{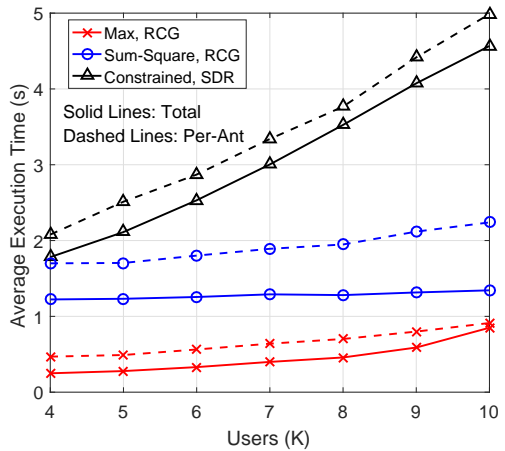}
    \caption{Average execution time for different optimizations, $\Gamma = 10\text{dB}$.}
    \label{fig:13}
\end{figure}
\begin{figure}
    \centering
    \includegraphics[width=0.85\columnwidth]{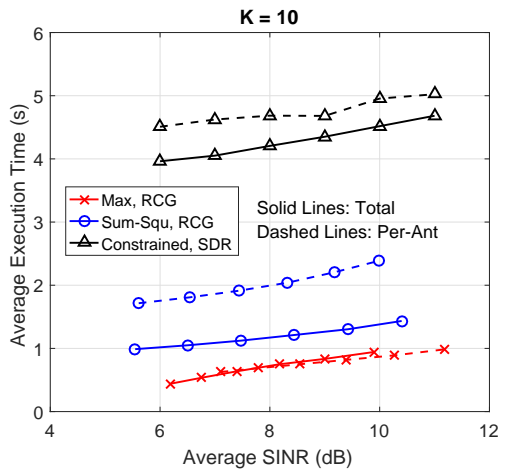}
    \caption{Performance-complexity trade-off for different optimizations,  $ K = 10$.}
    \label{fig:14}
\end{figure}
In Fig. 10, we explore the convergence speed of the manifold algorithms. We compare the value of cost functions for different weighted optimizations obtained in each iteration in conjunction with $K = 6, \Gamma = 10\text{dB}$, where two observations can be made: 1) For the optimizations with the same penalty terms, problems associated with total power constraints need fewer iterations to converge to the minimum than those with per-antenna power constraints; 2) For both power constraints, although the max SINR penalty and sum-square SINR penalty problems impose the same computational costs in each iteration, the former has a faster convergence rate, which suggests that the overall complexity of the max penalty problems is less than the problems with sum square penalty terms. In Fig. 11, we further characterize the overall complexity in terms of the average execution time for $\Gamma = 10\text{dB}$, since it is difficult to specify the analytic complexity when using the CVX. The simulations are performed on an Intel Core i7-4790 CPU 32GB RAM computer with 3.6GHz. The execution time here represents the time needed for solving the optimization for a single channel realization. It can be seen that all the weighted optimizations require much less time than the corresponding constrained problems, thanks to the low-complexity RCG algorithms. Quantitatively, the complexity is reduced to 50\% below for the scenarios considered, and the max penalty problem associated with total power constraint has the fastest convergence rate. Similar trends appear in Fig. 12, where the trade-off between execution time and the average SINR obtained have been shown for different beamforming methods with the same parameter configurations as in Fig. 7 and 8. Once again, we observe that the weighted optimizations provide up to an order of magnitude complexity reduction compared to the constrained optimizations, and the optimizations with max penalty terms achieve the best performance among all the beamforming methods.
\section{Conclusions}
A novel framework is proposed for the transmit beamforming of the joint RadCom system, where the beamforming schemes are designed to formulate an appropriate radar beampattern, while guaranteeing the SINR and power budget of the communication applications. We have considered both the separated radar and communications deployment, as well as the shared deployment. To further simplify the problem, we have considered weighted optimizations for the shared deployment that include the SINR constraints as the penalty terms in the objective function, which can be efficiently solved by manifold algorithms. Our numerical results show that the shared deployment has the far better performance than the separated deployment in terms of the trade-off between the quality of the beampattern and the downlink SINR. The proposed weighted optimizations obtain a similar performance to the original beamforming design at a much lower complexity, and therefore they offer a favorable performance-complexity trade-off.


%



\section*{Acknowledgment}
This research was supported by the Engineering and Physical Sciences Research Council (EPSRC) project EP/M014150/1 and the China Scholarship Council (CSC).

\ifCLASSOPTIONcaptionsoff
  \newpage
\fi



\bibliographystyle{IEEEtran}
\bibliography{IEEEabrv,Joint_Riem}

\end{document}